\documentclass[conference,compsoc]{IEEEtran}

\usepackage{tikz}
\usetikzlibrary{matrix,shapes,arrows,positioning,chains, calc}

\usepackage{amsmath,amsthm,amsfonts,amssymb}

\usepackage{tabularx} 
\usepackage{bbding}

\usepackage{makecell}
\usepackage{authblk}
\usepackage{xcolor}
\usepackage{listings}
\usepackage{booktabs}
\usepackage{multirow}
\usepackage{booktabs}
\usepackage{enumitem}
\usepackage{graphicx}
\usepackage{hyperref}
\usepackage{xcolor}
\usepackage{listings}
\usepackage{todonotes}
\usepackage{comment}
\usepackage{subcaption}
\usepackage{algorithm}
\usepackage{algorithmic}
\usepackage{hyperref}
\hypersetup{                    
  colorlinks,
  linkcolor={green!80!black},
  citecolor={red!70!black},
  urlcolor={blue!70!black}
}
\usepackage{xcolor,xspace}
\usepackage{url}

\usepackage{amssymb}
\usepackage{verbatimbox}
\usepackage{multirow}
\usepackage{scalerel}

\newcounter{index}

\newcommand{\ext}{\ensuremath{\mathtt{Ext}}\xspace}
\newcommand{\inj}{\ensuremath{\mathtt{Inj}}\xspace}
\newcommand{\edit}{\ensuremath{\mathtt{Edit}}\xspace}
\newcommand{\dec}{\ensuremath{{D}}\xspace}
\newcommand{\enc}{\ensuremath{{E}}\xspace}
\newcommand{\img}{\ensuremath{{x}}\xspace}

\newcommand{\el}{{\textit{et al.,} }}
\newcommand{\ie}{{\textit{i}.\textit{e}.,}\xspace}
\newcommand{\eg}{{\textit{e}.\textit{g}.,}\xspace}

\renewcommand{\el}{et al.\xspace}
\renewcommand{\ie}{i.e.,\xspace}
\renewcommand{\eg}{e.g.,\xspace}

\newcommand{\ipx}{\texttt{Instructpix2pix}\xspace}
\newcommand{\igc}{\texttt{Imagic}\xspace}
\newcommand{\invinj}{\ensuremath{\mathsf{RIW}}\xspace}

\newcommand{\ignore}[1]{}
\newcommand{\zztitle}[1]{\vspace{2pt}\noindent\textbf{#1.}}

\begin{document}

\title{A Somewhat Robust Image Watermark against Diffusion-based Editing Models}



\author[1]{Mingtian Tan}
\author[1]{Tianhao Wang}
\author[2]{Somesh Jha}
\affil[1]{University of Virginia}
\affil[2]{University of Wisconsin Madison}
\author{}
\pagestyle{plain}
\date{}

\maketitle

\begin{abstract}
Recently, diffusion models (DMs) have become the state-of-the-art method for image synthesis. Editing models based on DMs, known for their high fidelity and precision, have inadvertently introduced new challenges related to image copyright infringement and malicious editing. Our work is the first to formalize and address this issue. After assessing and attempting to enhance traditional image watermarking techniques, we recognized their limitations in this emerging context. In response, we develop a novel technique, \invinj (Robust Invisible Watermarking), to embed invisible watermarks leveraging adversarial example techniques. Our technique ensures a high extraction accuracy of $96\%$ for the invisible watermark after editing, compared to the $0\%$ offered by conventional methods. We provide access to our code\footnote{https://anonymous.4open.science/r/RIW-2F11/}. 
\end{abstract}


\section{Introduction}
\label{sec:introduction}
Recent years have seen remarkable contributions in computer vision through advancements in image generation, largely propelled by the advent of diffusion models (DMs)\cite{nonequilibrium}. This progress is partially due to their robust theoretical foundations\cite{ddpm,sde}, enhanced latent space inference~\cite{latent_space}, and efficient image-sampling techniques~\cite{ddim}. These factors have led DMs to commercial success, as seen in DALL-E~\cite{dalle} and Stable Diffusion~\cite{latent_space}. Additionally, DMs have spurred various emerging applications, highlighting their potential expansion in fields like data synthesis~\cite{dreambooth}, video synthesis~\cite{video_dm0,video_dm1,video_dm2}, text-to-3D synthesis~\cite{3d_dm0}, and image editing~\cite{ipx,imagic,glide,sdedit,blended,dual}.

Alongside the commercial success and potential applications of diffusion models, it also brings intellectual property (IP) concerns to the forefront~\cite{glaze}. Especially, the advancement of DM-based editing models~\cite{ipx,imagic,dalle-m,glide}, due to their hyper-realistic editing effects and fundamental differences from mere image generation tasks, still presents unresolved issues of unauthorized image use and deceptive edits.  

Several strategies to address these issues have been explored.  For example, Liang {et al.} have implemented adversarial training~\cite{pgd,fgsm} to curb unauthorized style transfer manipulation by DMs~\cite{mist-paper}. Similarly, Shan \el introduce adversarial noise into artworks training data to hinder DMs from learning and replicating unique artistic styles~\cite{glaze}. 
Several methods embed a custom watermark in training set~\cite{recipe,wdm_zhang} or during inference~\cite{diffusionShield,pips}, embedding an identifier in the produced images.
Notably, the recent work Tree-Ring~\cite{tree_ring} has achieved promising results by embedding watermarks~\cite{digital_water2002,digital_water2003} into the initial noise matrix of the diffusion process, thereby watermarking the model effectively. 

In this paper, we investigate image watermark schemes that establish image ownership post-editing. 
In this field, traditional methods, such as digital~\cite{digital_water2002,digital_water2003} or AutoEncoder~\cite{autoencoder_water,rivaGAN} based watermarks, are insufficient, especially with the advancement of DM-based image-editing models~\cite{evadingW}, because the changes introduced by modern image-editing models are unpredictable and prevalent, totally destroying the watermarks.

Our approach is to frame this as an optimization problem.  Taking any target image (e.g., an artwork) and a watermark (e.g., a random image), we embed them together so that the resulting image will maintain the semantic meaning of both images in the {\it latent space} of the editing model.  The underlying idea is to leverage the "mixup" effect~\cite{zhang2017mixup} where multiple images can co-exist in an image (i.e., the new image is a weighted average of multiple images), and the image-editing model can operate differently on different semantics, i.e., change the target image while leaving the watermark image untouched.  However, we cannot directly overlay the watermark image onto the target image, because this will make the modifications detectable. Our optimization ensures that the modifications are small.

Given the watermarked image, the next question is how to extract the watermark after it has been edited.  Similar to when we embed the watermark, the watermark in the edited image should also be unnoticeable.  Extracting watermarks while keeping them unnoticeable is challenging.  To enhance the extraction efficacy, we propose to boost the extraction signal by embedding multiple invisible watermarks at different positions within the image, and aggregating extracted signals from those positions. 

\begin{figure}[t]
    \centering
    \begin{subfigure}{0.5\textwidth}
    \includegraphics[width=.99\linewidth]{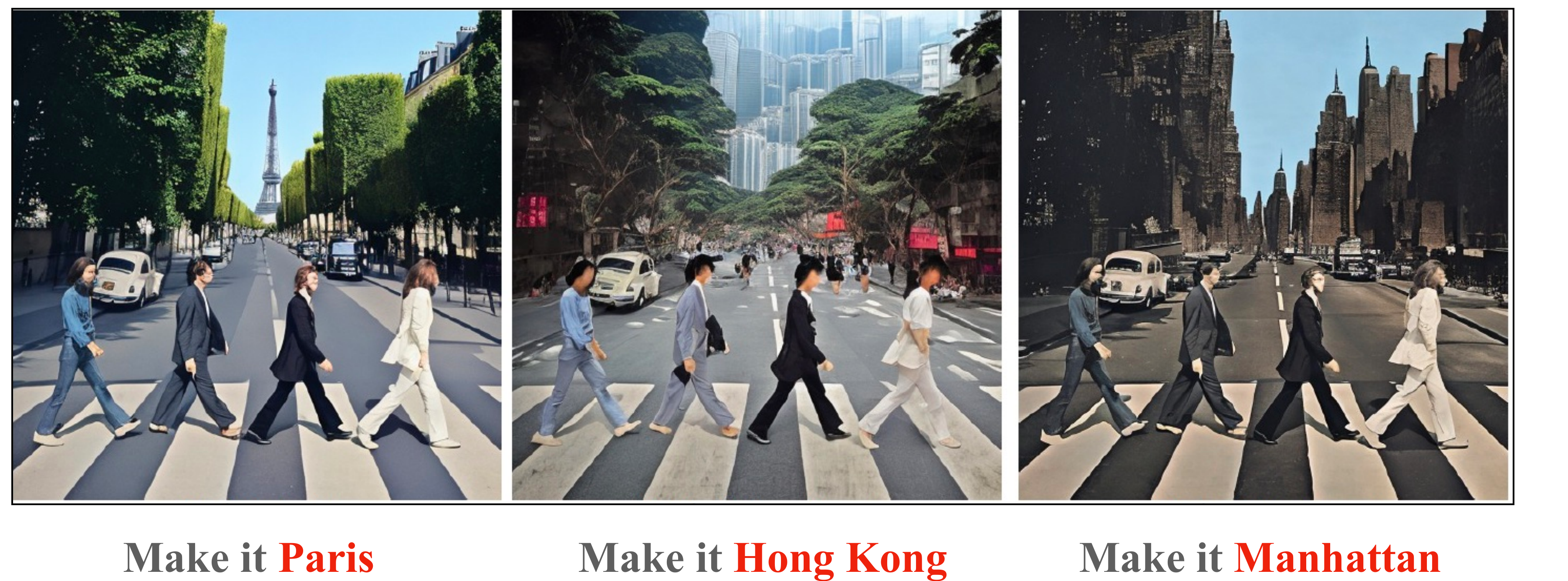}  \\
    \end{subfigure}
    \caption{For samples on editing model~\cite{ipx}, they vary based on the text prompt where modifications are required. For instance, ``Paris'', ``Hong Kong'', and ``Manhattan''.}
    \label{fig:ipx_pair}
\end{figure}

We have to limit the scope of our solution.  First, we cannot defend against arbitrary edits and only focus on minor edits (e.g., changing style, as shown in \autoref{fig:ipx_pair}).  We argue that even non-technically, it is hard to claim IP if the image is changed significantly.
Second, our method needs some knowledge about the diffusion model, but does not need access to the exact edit model or edit prompt.

Through an evaluation on two real-world datasets, the extended LAION-Aesthetics and TEDBench, as well as two state-of-the-art editing models~\cite{ipx,imagic} from different paradigms, we discovered that, compared to prior approaches~\cite{autoencoder_water,rivaGAN,digital_water2002,digital_water2003}, our watermarking technique demonstrates superior resistance against distortions introduced by editing models. For instance, post-editing by DMs, $96.0\%$ of our invisible watermarks can still be extracted, while traditional methods yield a $0\%$ success rate.

Our contributions are summarized as follows,
\begin{itemize}[leftmargin=*]
    \item We are the first to address the issues of image copyright and malicious modifications within editing models. Additionally, we formalize this problem and define our game framework.

    \item After assessing existing watermarking methods, we found them vulnerable to DM-based editing models. Despite our efforts using adversarial training and fine-tuning, current techniques remain inadequately prepared to tackle this new challenge.

    \item We propose an invisible watermarking technique, named \invinj (Robust Invisible Watermarking), based on visual-semantic effects. This ensures that our watermark remains intact even after editing, addressing the image copyright issues within the editing model framework.
\end{itemize}

The rest of this paper is structured as follows: 
In~\autoref{sec:related}, we review and discuss related works. 
In~\autoref{sec:preliminary}, we discuss the current editing models and the image copyright issues they introduce. We also define our threat model and discuss existing solutions. 
In~\autoref{sec:our_method}, we refine existing techniques and introduce our approach. \autoref{sec:setup} gives the evaluation setup and \autoref{sec:eval} provides detailed experimental results. 
We conclude in~\autoref{sec:conclusion}.

\section{Related Work}
\label{sec:related}
The application of semantic edits to authentic photographs or images has attracted considerable interest, particularly with the advent of advanced image synthesis techniques~\cite{goodfellow2014generative,nonequilibrium,ddpm}. Effective image editing demands that the model excel in image generation and interpret text-based editing instructions. Numerous studies have leveraged Generative Adversarial Networks (GANs)~\cite{goodfellow2014generative} for diverse image manipulations, such as domain translation between images~\cite{gan_translation_1,gan_translation_2,gan_translation_3} or transfer the image to a specific style~\cite{gan_transfer_1,gan_transfer_2,gan_transfer_3}. 
Meanwhile, with the rise of Diffusion Models (DMs)~\cite{nonequilibrium}, we have seen even better performance in image synthesis~\cite{synthesis_1,ddpm,class_condition,image_condition,song2019generative} as well as other multimodal tasks such as video~\cite{video_dm0,video_dm1,video_dm2} and audio~\cite{audio_0} generation.

With the advancement of DMs, concerns about image IP (Intellectual Property) have risen~\cite{nonequilibrium,ddpm,song2019generative,score_match,ptw}.  The realistic generative capabilities of DMs render it challenging for humans to distinguish whether images have been altered~\cite{ipx,imagic,dalle-m,glide,sdedit}. Consequently, this not only poses a threat to artwork IP but also facilitates the potential spread of misinformation through malicious image modifications.
To tackle this challenge, we categorize existing solutions into two approaches: (1) safeguarding artwork samples and (2) mitigating the misuse of diffusion models (DMs). 

\zztitle{Safeguarding Artworks}
To shield artworks from style mimicry through diffusion Text-to-Image models~\cite{latent_space,imagen}, Shan \el introduce GLAZE~\cite{glaze}. It adds noise to the original artwork so that DMs like Stable Diffusion~\cite{latent_space} and DALLE.E-m~\cite{dalle-m} cannot discern and adopt the unique art styles of these artworks.  Concurrently, Liang \el~\cite{mist-paper} launched the MIST project~\cite{mist-pj}, utilizing adversarial training to incorporate noise into artworks. When these altered images are shared online, adversaries cannot obtain a high-quality image after the style transfer.  Chen \el~\cite{chen2023editshield} similarly add small perturbations to images to disrupt modifications.

\zztitle{Mitigating Misuse of DMs}
Zhao \el~\cite{recipe} propose to introduce watermarks into training samples, i.e., artworks, utilizing an AutoEncoder model~\cite{autoencoder_water}. Once trained on these watermarked samples, DMs consequently produce images bearing imperceptible watermarks. These watermarks can be subsequently retrieved through the decoder, establishing the ownership of public DMs. Similarly, Liu \el employ a novel technique to watermark the latent DM~\cite{wdm_zhang}, training their model on paired data of watermarked prompts and watermarking images. Consequently, the prompt serves as a trigger to generate their watermarked image, verifying model ownership. Contrarily, certain research takes the route of introducing watermark directly into the diffusion process~\cite{pips,tree_ring,diffusionShield}, successfully extracting the watermark and leveraging the determinism inherent to the DDIM~\cite{ddim} generation mechanism.



\section{Preliminaries} 
\label{sec:preliminary}
In this paper, we address the abovementioned issue (protecting image IP in editing models) from {\it a new angle using image watermarking schemes}.  Specifically, we aim to add watermarks to images so that after editing, we can detect that these edited images are from those images.
If adversaries malevolently alter artwork styles, infringe copyrights, or manipulate the semantic content of original images using image editing models, these watermarks can help claim ownership.

In what follows, we begin by reviewing image editing models, and then formally describe our threat model, and review existing image watermarking schemes.


\subsection{Image Editing}


We focus on diffusion-based image editing models~\cite{sdedit,blended,dalle,dreambooth}, which show remarkable results due to the outstanding performance of Denoising Diffusion Probabilistic Models (DDPM)~\cite{ddpm} (details deferred to~\hyperref[subsec:dms_edit]{Appendix~\ref*{subsec:dms_edit}}). Contrary to traditional GAN-based editing models such as StyleGAN~\cite{karras2019style}, which are limited to singular editing functionalities (\eg a model solely dedicated to converting facial images into smiling faces), recent DMs-based image editing models enable zero-shot editing of images merely with text instructions and achieve a superior editing quality~\cite{ipx,imagic}. 

Specifically, these models typically employ an encoder \enc to map images (denoted by \img) into a latent space, while simultaneously mapping text edit prompts (denoted by $p$) into an embedding, often using tools like CLIP~\cite{clip}, representing by $C$, due to its proficient textual understanding capabilities. 
Subsequently, a decoder \dec is used to reconstruct the high-resolution image, fusing the image and text information:
\begin{align}
      \img_{e}=\dec\left [ \enc(\img),C(p)  \right ] \label{eq:edit}
\end{align}
or more abstractly, $\img_{e}=\edit(\img, p).$

\subsection{Problem Statement}

We assume there is Alice who has access to an image-editing model, denoted by $\edit$, and uses it to edit images, with text prompt $p$.  We use a subscript $e$ to denote edition: an image \img is edited into ${\img}_e\gets\edit(\img,p)$. 
It is worth noting that the edition will be focused on modifying the style or changing minor things.  For example, in~\autoref{fig:ipx_pair}, the streets are changed, while the main theme stays the same.  We argue that if the change can be arbitrary and the change is too much such that the whole image is changed, there is no IP issue anymore.

On the defense side, Bob embeds invisible watermarks into the images before they are published and before they are accessed by the attacker.  We use $\hat{\img}\gets \inj(\img, w)$ to denote the injection of a watermark $w$ into image \img. 
Note that Bob may have some high-level ideas about the architecture of the editing model $\edit$, but does not know the details of $\edit$, and neither can he modify $\edit$.  Moreover, Bob has no idea what the image-editing prompt $p$ Alice will use.
After modifications by Alice, Bob can tell, and, more importantly, prove this edited watermarked image $\hat{\img}_e$ comes from the original image \img.  \autoref{tab:notations} gives the key operations.


\subsubsection{Formalizing Desired Goals} 
We identify two goals in the process, which we formalize as follows.

\zztitle{Invisibility}
First, the injected watermark should be invisible.  Moreover, once edit models edit the watermarked images, Alice should not find out the image has been watermarked.

\zztitle{Extractability}
The watermark should be robust and extractable on any edit operations (i.e., any edit model and prompt).  As mentioned earlier, the modification cannot be large; otherwise, it is unclear if there are IP issues anymore.  We will quantify the amount of modification later.

\begin{table}[t]
    \caption{Summary of key operations.}
    \label{tab:notations}
    \resizebox{0.48\textwidth}{!}{
    \centering
    \begin{tabular}{cc|l}
        \toprule
        \textbf{Party}&\textbf{Operations}&\textbf{Description}\\
        \midrule

        Bob&$\hat{\img}\gets \inj(\img, w)$&Inject watermark $w$ into image \img \\
                
        Alice&$\hat{\img}_e\gets \edit(\hat{\img}, p)$&Edit image $\hat{\img}$ with text prompt $p$\\

        Bob&$w'\gets\ext(\hat{\img}_e)$&Extract the watermark\\  
        

        
        

        
        \bottomrule
    \end{tabular}
    }
\end{table}

\begin{figure}[t]
\centering
\begin{tikzpicture}
\matrix (m)[matrix of nodes, column  sep=0.5cm,row  sep=2mm, nodes={draw=none, anchor=center,text depth=0pt} ]{
    Alice                                    &                      & Bob\\
                                             &                       &  $\img^0\xleftarrow{\$}\mathbb{X}$ \\[-3mm]
                                             &                       &  $\img^1\gets\inj(\img^0,w)$ \\[-4mm]
                                         & $b\xleftarrow{\$}\{0,1\}$ & \\[-3mm]
                                         & Send $\img^b$               & \\[-1mm]
$\img^b_e\gets\edit(\img^b,p)$                   &                   &\\[-4mm]
                    &    Send $\img^b_e$      & \\[-4mm]
                       &         & $w'\gets\ext(\img^b_e)$\\[-3mm]
 Guess $b_a$                         &         & Guess $b_b$\\
};
\draw[shorten <=-.25cm,shorten >=-.25cm] (m-1-1.south east)--(m-1-1.south west);
\draw[shorten <=-.25cm,shorten >=-.25cm] (m-1-3.south east)--(m-1-3.south west);
\draw[shorten <=-.25cm,shorten >=-.25cm,-latex] (m-5-2.south east)--(m-5-2.south west);
\draw[shorten <=-.25cm,shorten >=-.25cm,-latex] (m-7-2.south west)--(m-7-2.south east);
\end{tikzpicture}
\caption{We formalize the problem definition into a game between Alice and Bob. Alice wins if $b_a=b$ (Alice can detect whether there is a watermark based on $x^b$ and $x_e^b$), and Bob wins if $b_b=b$ (Bob can extract the watermark after edits).}
\label{fig:game}
\end{figure}
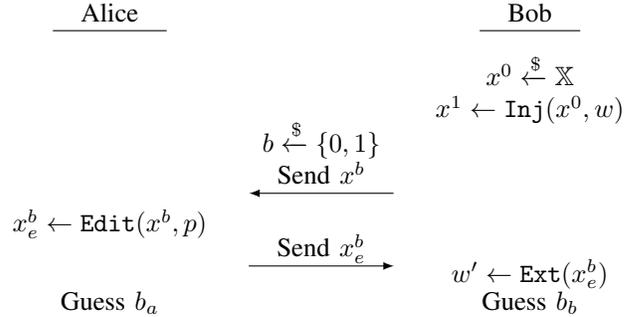

To capture these goals more precisely, we define a crypto-style game between Alice and Bob in~\autoref{fig:game}.  First, Bob gets a randomly sampled image from a pre-defined set, and inserts a watermark $w$ into it.  The two images are denoted by $\img^0$ and $\img^1$, respectively.  Then one of the images $\img^b$, where $b$ is a random bit, is sent to Alice for editing.  Then based on the edited image $\img^b_e$, Bob tries to extract the inserted watermark.  Ideally, $w'\approx w$ only if $b=1$, and in which case, Bob guesses $b_b=1$; if $w'$ is random or is different from $w$ Bob guesses $b_b=0$.  Bob wins if $\hat{b}=b$.  At the same time, we want the watermark to be invisible.  So if Alice can distinguish if the image was watermarked or not, Alice can win.  Both Alice and Bob can win at the same time, but our goal is to make only Bob win. 

Our game operates in the black-box setting, i.e., Alice does not know the details about the \inj and \ext operations.  On the other hand, Bob has some knowledge about the \edit operation (e.g., some publicly available edit model).  We acknowledge this is a limitation of our study, and leave the exploration of the more challenging scenarios (e.g., Alice knows \inj and \ext and becomes adaptive when choosing prompts) as future work.

\subsubsection{Problem Scope} 
As mentioned in related work, there has been a surge of research on knowledge copyright protection within diffusion models, categorized into {\it preventing} image styles from being learned by diffusion models~\cite{glaze}, guarding against style transfers~\cite{mist-paper}, and {\it safeguarding diffusion models} from unauthorized theft or misuse~\cite{diffusionShield,tree_ring,recipe}. Our watermarking technology primarily focuses on {\it detecting (via watermarks)} unauthorized modifications or malicious edits of {\it artwork or photographs} by diffusion model-based editing tools. Artists and photographers can utilize our tool to embed subtle, nearly undetectable perturbations into their work, ensuring minimal impact on the original image quality before online publication. 

The goal of our watermark is to be invisible and not to be robust against adaptive modifications:
adaptive editing attack~\cite{adaptive} or image reconstruction based on generative model~\cite{zhao2023invisible} might remove our watermark. 
Also, while the game definition is flexible, in our experiment, we limit the text prompt $p$ so that only minor changes will be made to image $x$.  We argue that deciding whether greater change violates IP protections is a broader problem and out of the scope of this paper.
\subsection{Existing Image Watermarking Schemes}
\label{subsec:solu}


\subsubsection{Digital Image Watermark}


Early work on image watermarking focuses on designing digital watermarks~\cite{digital_water2002,digital_water2003} that
integrate watermarks into colored images.  Naive methods insert watermarking messages into specific locations of the image; more recent work 
embeds the watermark in the transformed space to make it subtle and less discernible to the human eye, but can be extracted when necessary using the inverse transformations.




\begin{figure}[t]
    \centering
    \begin{subfigure}{0.50\textwidth}
    \includegraphics[width=.95\linewidth]{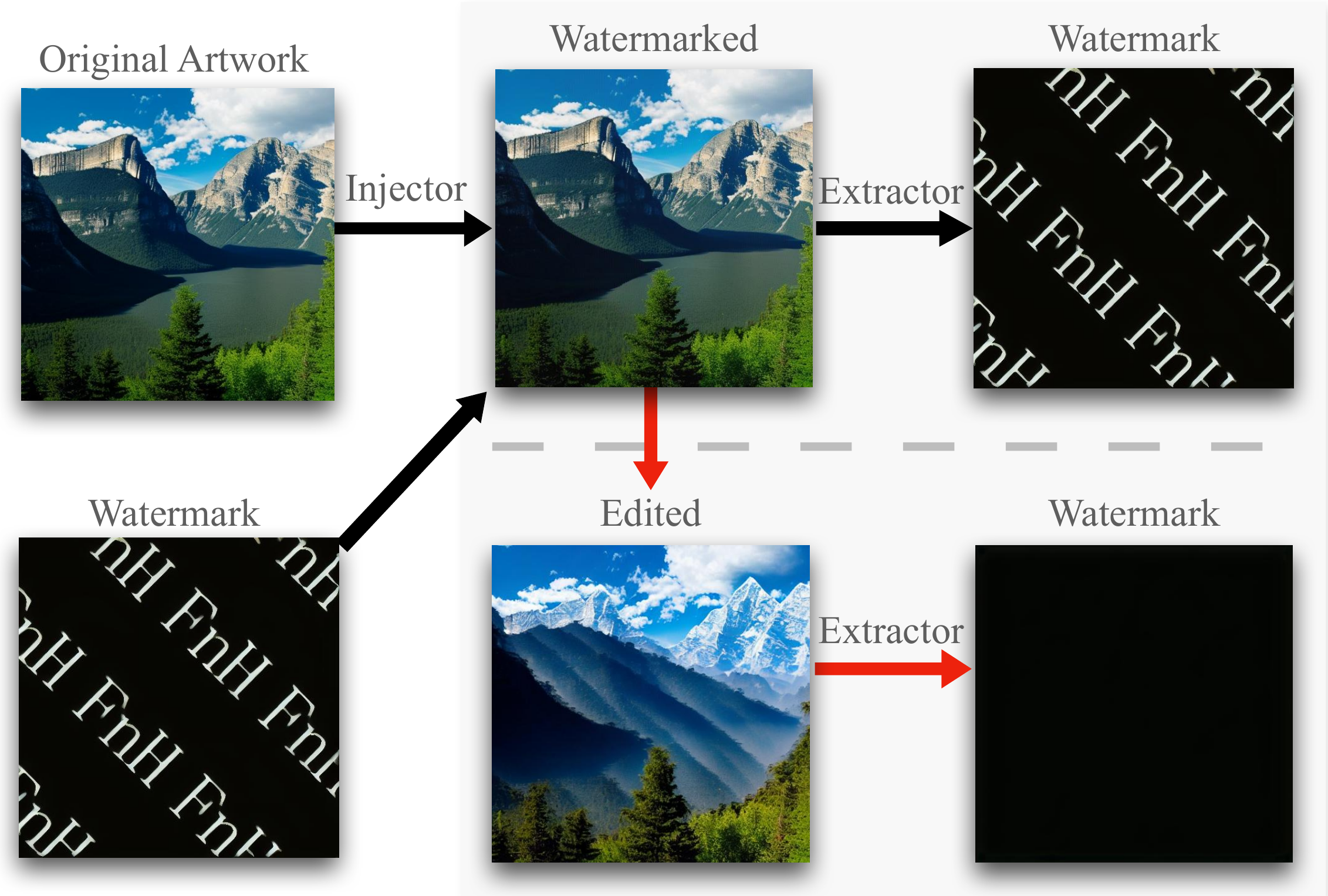}
    \end{subfigure}
    \caption{The example of watermarking using an AutoEncoder. The original artwork, after being watermarked by Injector \inj, can have the corresponding watermark successfully extracted by Extractor \ext. After being edited by \ipx with the text prompt ``Have it be the valley of the {Ten Peaks} in the {Himalayas}'', \ext can no longer retrieve the watermark.}
    \label{fig:ae_example}
\end{figure}

\subsubsection{AutoEncoder-based Watermark}
\label{subsec:autoencoder}
More recently, researchers built watermarking schemes using the AutoEncoder architecture~\cite{adaptive,recipe}. In the AutoEncoder framework~\cite{autoencoder_water,rivaGAN}, an encoder is trained to embed watermarks, while a decoder is trained for watermark extraction. Note that the encoder and decoder have similar functionalities as those defined in \autoref{eq:edit}, and we slightly abuse notations and reuse \enc and \dec to denote them.  But the internal neural network architectures are quite different.


Specifically, an encoder \enc is trained as \inj to embed watermarks, while a decoder \dec is trained as \ext for watermark extraction. Specifically, during the embedding phase, both the image \img and watermark $w$ are fed into the \enc and encoded into a latent space. In this latent space, the encoder integrates the watermark with the image and then upsamples it back to the image domain, yielding a subtly modified image $\hat{\img}\gets \enc(\img,w)$ that remains visually indistinguishable from the original format. 
The decoder \dec is then trained to extract the watermark $w$ from $\hat{\img}$. The optimization of the AutoEncoder is as follows, 
\begin{align}
     \underset{\dec,\enc }{\mathrm{arg\,min}} \; \left \|  \dec(\hat{\img})-w \right \|_{\ell_2} + 
     \lambda[\left \| \hat{\img} - \img \right \|_{\ell_2} + L_{lpips}(\hat{\img},\img)]
     \label{eq:original_obj}
\end{align}
where the first term ensures the decoder can extract the watermark $w$ embedded by the encoder, and the second term guarantees that the watermarked image $\hat{\img}$ visually approximates the original \img (by measuring similarity between $\hat{\img}$ and \img using $\ell_2$ norm and LPIPS loss~\cite{lpips}; but other metrics will also apply), and $\lambda$ is used to balance the scales of these two terms. 

However, our evaluations (details in~\autoref{subsec:compare}) reveal that images watermarked using this method struggle to withstand distortions introduced by the edit models. As a result, it becomes challenging to extract the embedded watermark from the modified image after editing. 
As illustrated in~\autoref{fig:ae_example}, the watermarked image $\hat{\img}$ is visually indistinguishable from the original image, and the decoder \dec successfully extracts the embedded watermark $w$ from the $\hat{\img}$. However, post-editing by image-editing models, the watermark extracted from image $\hat{\img}_{e}$ is difficult to see. 

\section{Our Method}
\label{sec:our_method}
We identify two fundamental flaws for the existing AutoEncoder-based watermarks: (1) they do not consider image edits, and (2) the `one-model-fit-all' approach inherently makes training such a model more challenges, i.e., because the edits are so diverse, it is difficult to ensure the training converge.
Overcoming the first issue is easier: we just need to redefine the objective function to incorporate image edits.  On the other hand, the second issue is more fundamental, and we propose to decouple the encoding and decoding phases and consider/optimize them separately.

\subsection{Warmup: Incorporating Image Edit}
Given that the AutoEncoder-based solution is not robust against perturbations from image-editing models, we make two efforts to improve it.  First, we incorporate adversarial learning~\cite{adv_learning} into the optimization objective as follows:
\begin{align*}
     \underset{\dec,\enc }{\mathrm{arg\,min}} \;  \left \|  \dec(\hat{\img}+ \alpha {N})-w \right \|_{\ell_2} + 
     \lambda[\left \| \hat{\img} - \img \right \|_{\ell_2} + L_{lpips}(\hat{\img},\img)]
\end{align*}
Compared to the original objective, here we introduce an isotropic Gaussian noise denoted by ${N}$, and $\alpha$ represents the strength of the noise. Now \dec is able to extract the watermark $w$ from images with $\alpha$ scaled Gaussian noise, but it still fails to extract the watermark after the edited models. This is because the Gaussian noise we introduced and the diffusion process's perturbation on the image are not exactly the same. 

To address this, we further fine-tuned \dec using watermarked images that had been edited, and the objective function becomes
\begin{align}
     \underset{\dec,\enc }{\mathrm{arg\,min}} \;  \left \|\dec(\hat{\img}_{e}),w \right \|_{\ell_2} +
     \lambda[\left \| \hat{\img} - \img \right \|_{\ell_2} + L_{lpips}(\hat{\img},\img)]\label{eq:autoencoder-optimized}
\end{align}
Unfortunately, in our evaluation, \dec fails to fit edited images during the training process, and the training loss escalates $0\%$. We hypothesize that the watermark information added by \dec is destroyed by the editing process. 

\subsection{Watermark Injection}
\label{subsec:iij}
Having identified the difficulty of training the autoencoder, our approach decouples the watermark injection and extraction processes and consider them separately.
One significant distinction from prior solutions is we formulate this 
process as a {\it per-image} optimization problem.  Here the challenge is that different from training a model explicitly, we need to identify the direction of the optimization.  A natural consideration is to make the watermark invisible, but then how to make sure it will be extracted later (without knowing what the extractor does)? And how to ensure the optimization is efficient?

To tackle the first question, we propose to make the watermarked image $\hat{\img}$ carries information about the watermark $w$ in the {\it latent/embedding space}.  That is, suppose we know the encoder $\enc$ in the image-editing model (i.e., $\enc$ from \autoref{eq:edit}), we can ensure $\enc(\hat{\img})$ maintains information about $\enc(w)$, or equivalently, the distance between $\enc(\hat{\img})$ and $\enc(\img + \alpha\cdot w)$ is small, where $\alpha$ denotes the strength of the watermark. The intuition behind working in the latent space is that image editing models typically target and modify only the instruction part, while retaining unrelated semantic information, thus retaining the watermark.

At the same time, we want to make sure $\hat{\img}$ is similar to $\img$, as quantified earlier.  We borrow the similarity measurement from previous optimization objectives, but with the LPIPS loss replaced by a post-edition invisibility penalty.  Specifically, we borrow the decoder \dec from Stable Diffusion~\cite{latent_space} ($\enc, \dec$ is a pair; note that although we also use these two notations to denote the encoder and decoder of the autoencoder, their internal architectures are significantly different), which is utilized to retrieve the generated image from the latent space.  $\dec(\enc(\hat{\img}))$ gives the edited watermarked image, but without any prompt.  Intuitively, if $\dec(\enc(\hat{\img}))$ is similar to $\img$, meaning that the watermark is invisible, then after the actual edition, the watermark will also be invisible.
This leads to our proposed optimization objective:
\begin{align}
    \underset{\hat{\img} }{\mathrm{arg\,min}} & \; \left \| \enc(\hat{\img})-\enc(\img + \alpha\cdot w)\right\|_{\ell_1}  \notag\\
    &+\lambda\left[\left \| \hat{\img}-\img\right\|_{\ell_2} 
    + \left \| \dec(\enc(\hat{\img}))-\img \right \|_{\ell_2}\right]  \label{eq:our_obj} 
\end{align}
Here the first component compares in the embedding space, and $\ell_1$ norm is a natural metric; but like before, other metrics can also be used.
We explore the interplay between $\alpha$ and watermark detection precision in~\autoref{subsec:strength}.  {The parameter $\lambda$ serves as a weighting factor to balance the strength of the watermark in the {latent space} with the quality of image generation.} Note that we employ an encoder in our optimization function similar to the target editing models. Our results indicate that this optimization approach is not relying on having identical encoders in terms of training parameters and data. By adjusting parameters $\alpha$ and encoder loss $\lambda$, i.e., {latent space} component in the optimization objective, we can modulate the watermark's intensity post-editing, rendering it invisible, as depicted in~\autoref{fig:ipx_example} and~\autoref{fig:imagic_exam}.




 

To ensure the optimization problem is solved efficiently, we leverage and adopt the PGD~\cite{pgd} algorithm, which was proposed to find adversarial examples and is efficient.
Our optimization process is given in~\autoref{alg:iij_alg}. Given $\img$ and $w$, we first obtain $\hat{\img}^0$, and execute a series of optimization steps, in a way similar to PGD: in each step, take the sign of the gradients multiplied by a scale (intuitively a `learning rate') of $\mu$ and the change, and limit the overall change from the original sample $x$ by $\varepsilon$, and update. 
\begin{algorithm}[t]
    \caption{Watermark Injection Pipeline}
    \label{alg:iij_alg}
    \begin{algorithmic}[1]
    \REQUIRE{
        Image \img, transparency $\alpha$, watermark $w$, encoder \enc, decoder \dec, optimization rate $\mu$, upper bound $\varepsilon$, and number of steps $T$.  
        }
    \STATE {Define objective function $L$ based on \autoref{eq:our_obj}} 
    \STATE {Generate target image $\img'\gets \img+\alpha\cdot w$} 
    \STATE {Initial watermarked sample $\hat{\img}^{0}\gets \img'$}
    \FOR{$t \gets 0$ to $T-1$}
    \STATE { $\hat{\img}^{t} \gets  \hat{\img}^t + \mu\cdot sgn(\bigtriangledown_{\hat{\img}^t}L(\enc,\dec,\hat{\img}^t,\img' )) $ }
    \STATE { Calculate $\delta=\hat{\img}^{t}-\img$ and bound it with $\varepsilon$ in $\ell_1$ }
    \STATE { Update $\hat{\img}^{t+1}$ by $\hat{\img}^t + \delta $ }
    \ENDFOR
    \RETURN {$\hat{\img}^T$\;}
    \end{algorithmic}
\end{algorithm}

\subsection{Watermark Extraction}
\label{subsec:method}
To extract the watermark, a baseline solution is to directly calculate the pixel-wise distance between the watermark $w$ and the edited image $\hat{\img}_e$, and can claim successful extraction if the distance is below a threshold.
However, we found the pixel-wise measurement is not stable in edit models, because the editions, while semantically minor, can cause a lot of pixels to be modified, potentially leading to a large pixel-wise distortion in the extracted watermark.  Moreover, as our optimization goal in~\autoref{eq:our_obj} involves making the watermark invisible post-edition, this intuitively cannot work.


To overcome this issue, we propose to extract at a higher level (semantic level).  Here, we propose a simple yet effective design, backed by two intuitions: (1) focus on image semantics instead of details/pixels, (2) boost accuracy by `divide and conquer'.

Specifically, we use simple images that are just plain text with black backgrounds as watermarks (and to differentiate, we call them text watermarks and the previous general watermark as image watermarks). {For examples of text and image watermarks, refer to~\autoref{fig:image_w} and~\autoref{fig:ipx_igc_examples}, respectively.} This ensures the higher-level semantics of images are easier to capture (i.e., just recognize texts in the image, using off-the-shelf OCR models).  We also segment the watermark into $K$ regions, so that we will have more confidence if we can recognize texts (which can also serve as the passcode of the watermark) in most regions correctly.

\subsection{Discussion}
\label{subsec:discuss}

To summarize, although we express our target similarly to the optimization objectives in the AutoEncoder-based solutions (for ease of understanding), the underlying technique is fundamentally different: 
(1) We decouple the watermark injection and extraction processes and consider them separately.
(2) We formulate the watermark injection process as a {\it per-image} optimization problem using PGD.
(3) We leverage information about the encoder and decoder and rewrite the objective to incorporate different possibilities of image edition.

The efficacy of our watermarking method is attributed to their shared foundation on Stable Diffusion~\cite{latent_space}. Although the encoder we use during watermark optimization is not identical to those used in the edit models, these encoders share similarities.  Especially, our watermark's resilience to distortions in the editing process is attributed to integrating the edit model's latent space information during optimization, allowing it to convey visual semantic information similar to that of the target image in the latent space. Another potential reason is: edit models strive to preserve the original image's semantic essence and only modify specific instruction-driven sections. 
We argue that due to the widespread open-source availability of generative models, such as Stable Diffusion~\cite{latent_space}, accessing similar versions of encoders is not challenging. 
Although our watermark, based on visual semantic information, exhibits a higher preservation rate in the edit model, our goal is not to defend against adaptive attacks: adaptive editing attack~\cite{adaptive} or image reconstruction based on generative model~\cite{zhao2023invisible} might remove our watermark. 
We defer the exploration of such attacks to future work. Moreover, since our watermark must be imperceptible to the human eye, yet present within the latent space of the diffusion, the model optimization of $\hat{\img}$ relies on the {\it latent space} information of the edit model (we rely on the transferability of models and assume white-box access to an older version of the edit model in the experiment). This leads to the challenge of generating high-quality watermarks in a complete black-box scenario.

\section{Experiment Setup}
\label{sec:setup}
\zztitle{Datasets} We evaluate our approach using two datasets: the extended LAION-Aesthetics and TedBench. The LAION-Aesthetics has been widely used for training diffusion models~\cite{dalle,ipx,latent_space}, while TedBench~\cite{imagic} has been introduced as a benchmark dataset to measure edit models. 
\begin{itemize}[leftmargin=*]
\item LAION-Aesthetics\footnote{http://instruct-pix2pix.eecs.berkeley.edu/}: Due to its large size and diverse content, the primary use of this dataset is image generation. We apply LAION-Aesthetics V2 6.5+~\cite{schuhmann2022laion}, a subset of the LAION's 5B samples, which comprises 625K image-text pairs and includes various image types, such as paintings, digital artwork, and photographs. Brooks \el expand the dataset to include pairs of images before and after editing, along with their corresponding edit instructions by fine-tuning GPT-3~\cite{gpt3} and employing the Prompt-to-Prompt~\cite{hertz2022prompt} technique. Additionally, they apply a CLIP-based metric~\cite{ipx} to filter out noise like nonsensical or undescriptive captions from this dataset, enhancing its quality and applicability.

\item TEdBench\footnote{https://github.com/imagic-editing/imagic-editing.github.io/}: Kawar \el~\cite{imagic} have proposed TEdBench (Textual Editing Benchmark), a benchmarking dataset comprising 100 pairs of input images and target text that describe intricate non-rigid edits. 
\end{itemize}

\zztitle{Image Editing Model} We utilize two state-of-the-art image editing models, \ie \ipx~\cite{ipx} and \igc~\cite{imagic}. The design principles of these two models are entirely different: One is built as an editing model by training on a pair of images--one representing the subject to be edited and the other representing the image post-editing~\cite{ipx}--while the other is based on fine-tuning the model using images that require editing~\cite{imagic}. 
Both models harness the capabilities of latent space diffusion models~\cite{latent_space} as their foundational generative framework.  Detailed descriptions about these models are deferred to~\hyperref[subsec:edit_models]{Appendix~\ref*{subsec:edit_models}}. Additionally, the editing prompt styles for these two models are quite distinct. In \ipx, the edit instruction typically begins with a verb, serving as a directive, while in \igc, it fully describes the post-editing image, thus guiding the model’s editing process. However, the dataset we used for evaluation, such as TEDBench~\cite{imagic}, does not encompass both of these prompt styles. To address this, we employed the GPT-4 API to automatically transfer and generate a variety of text prompt styles, facilitating compatibility with both \ipx and \igc model. More details refer to {\it Text Edit Prompt} in~\hyperref[subsec:edit_models]{Appendix~\ref*{subsec:edit_models}}.

\zztitle{Optimization Details}
In our optimization function, we utilize the version ``sd-v1-4.ckpt'' for the encoder, while the edit model \ipx employs ``sd-v1.5.ckpt'' and \igc uses the ``sd-v1-4-full-ema.ckpt'' version. Due to its fine-tuning paradigm, this model might slightly deviate from the original one. 
We employ the LPIPS (Learned Perceptual Image Patch Similarity) loss~\cite{lpips} to ensure the fidelity of our watermarked image $\hat{\img}$ in the second term of the optimization objective. 
All results presented in this paper are based on settings $\varepsilon=12/255$, $\mu=2/255$. We set $\lambda=1$  in the evaluation of modification $\alpha$, and set $\alpha=0.4$ in the evaluation of modification $\lambda$. Appropriate adjustments of $\lambda$ may lead to improved watermark quality, and we will explore these in future work if needed.  
 
\begin{figure*}[th!]
    \centering
    \begin{subfigure}{0.99\textwidth}
        \centering
        \includegraphics[width=.9\linewidth]{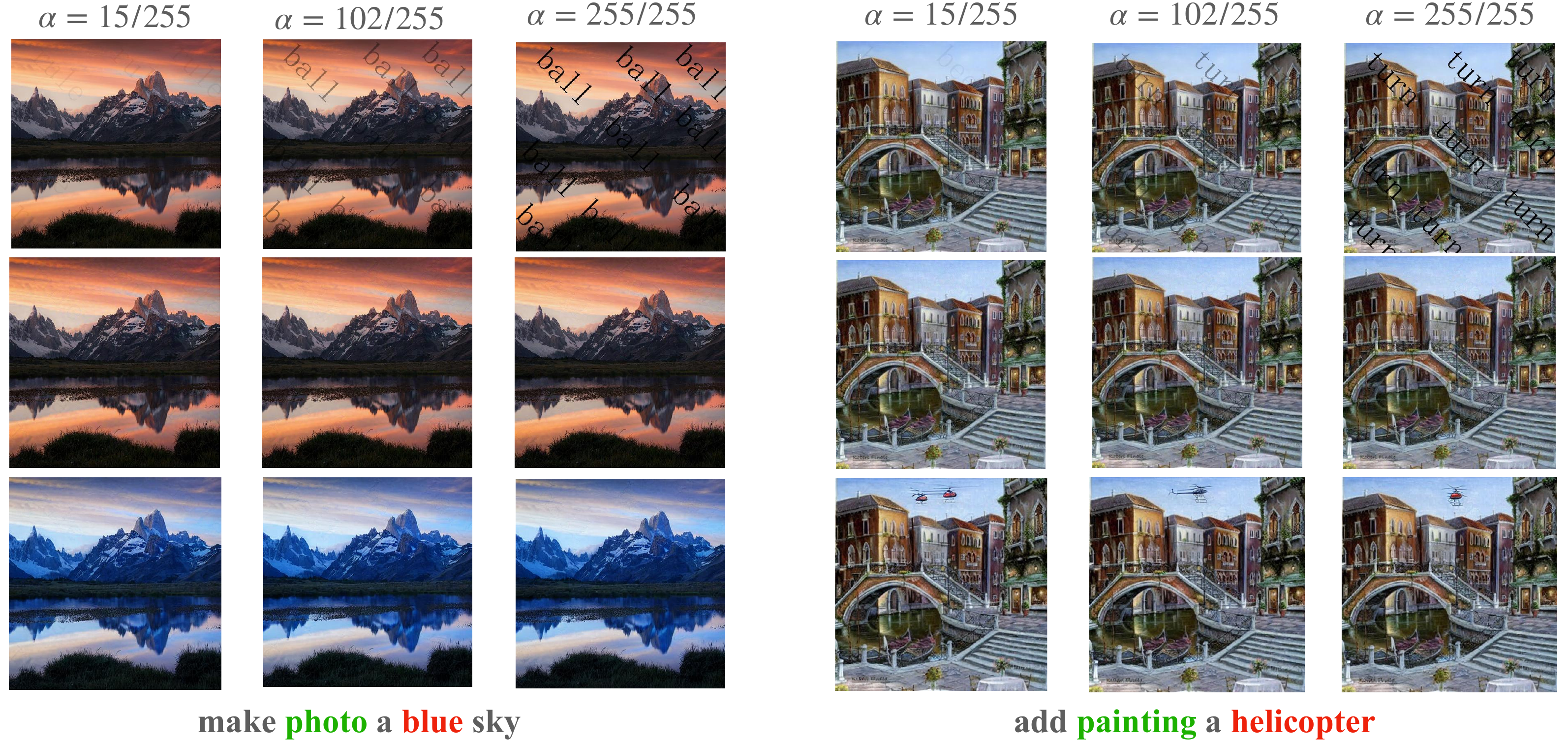}  
        \caption{Examples of the \ipx model.}\label{fig:ipx_example}
    \end{subfigure}
    \begin{subfigure}{0.99\textwidth}
        \centering
        \includegraphics[width=.9\linewidth]{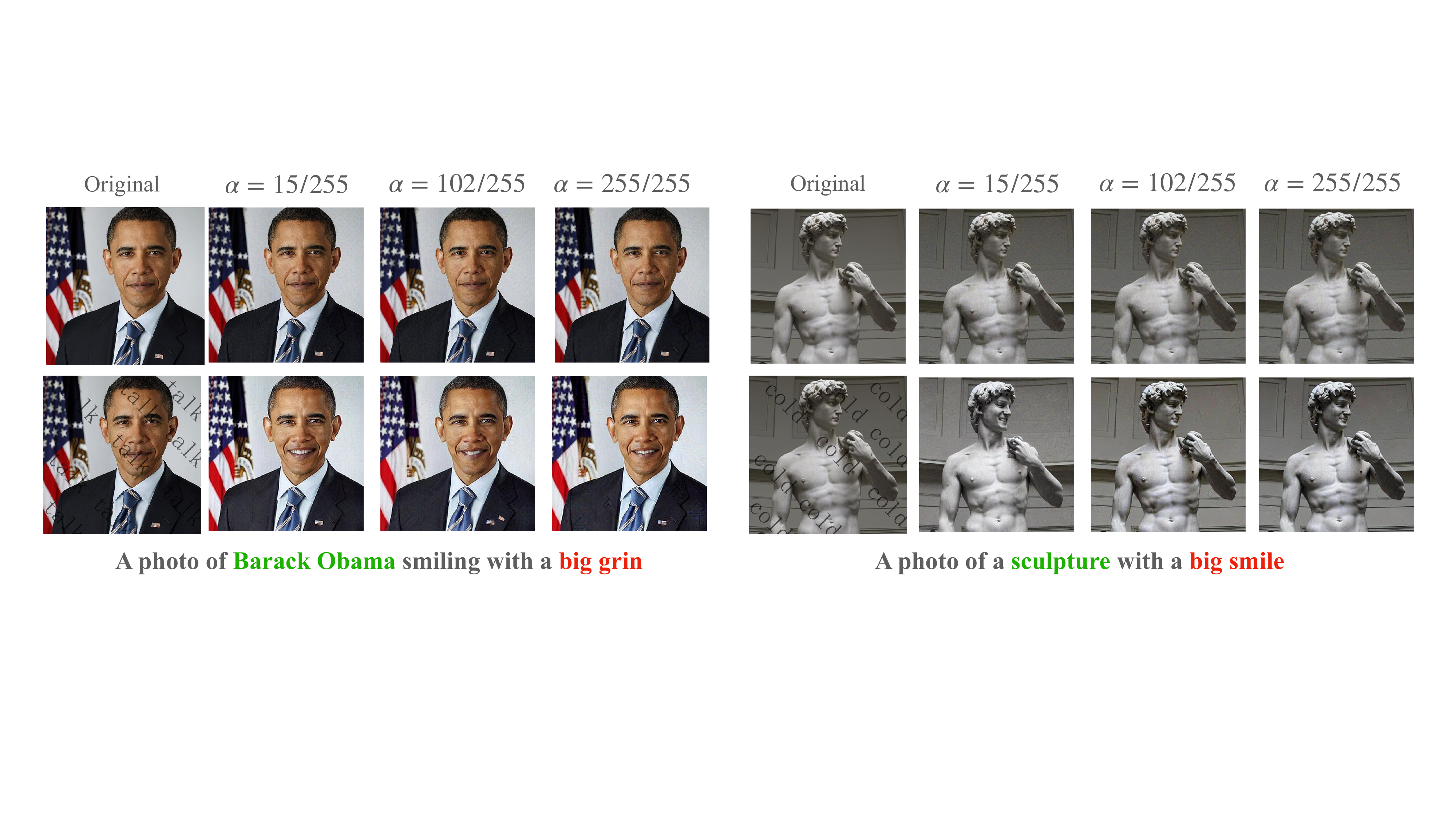}  
        \caption{Examples of the \igc model.}\label{fig:imagic_exam}
    \end{subfigure}
    \caption{In both subfigures, the first row represents the target images, denoted as $\img'$, exhibiting varying watermark transparency $\alpha$. The second row displays the images after the watermark is injected, denoted as $\hat{\img}$.  The last row illustrates the corresponding edited images, denoted as $\hat{\img}_e$. 
    }
    \label{fig:ipx_igc_examples}
\end{figure*}

\zztitle{Watermarks Used}
We crop the image into $9$ segments/blocks, with each size $(173,66)$, corresponding to the nine individual watermarks $w$ we have embedded, each consisting of four letters as seen the first example in \autoref{fig:imagic_exam}.

\begin{figure}[t]
    \centering
    \hspace*{0.2cm}
    \begin{subfigure}{0.5\textwidth}
    \includegraphics[width=8cm]{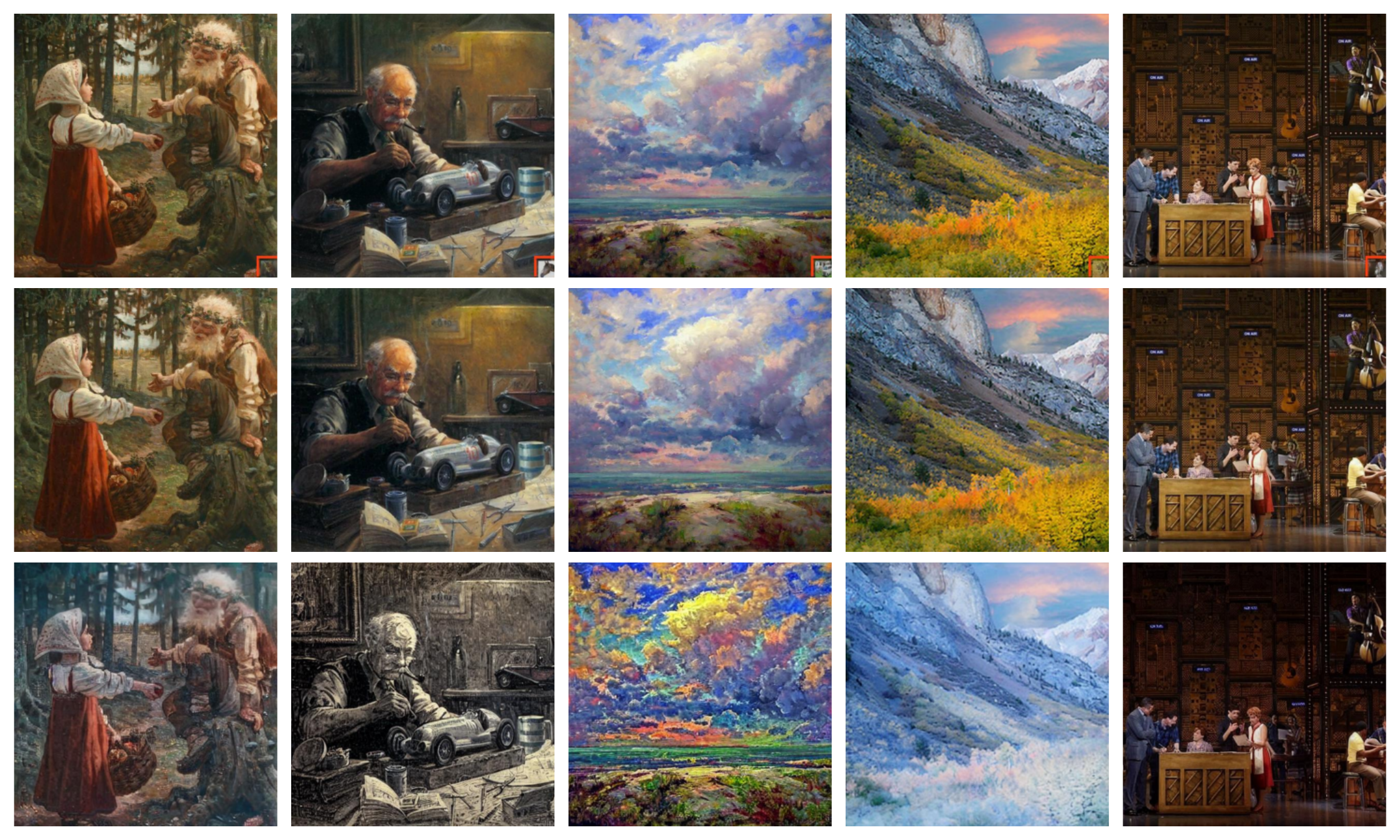}  \\
    \end{subfigure}
    \caption{{The first row displays the target image overlaid with an image watermark, featuring a $32\times 32$ sized watermark in the bottom right corner (marked with a red box), randomly selected from CIFAR10~\cite{cifar}. The second row showcases the invisible watermark added by our technique, and the third row illustrates the outcome post-editing of the watermarked image.}}
    \label{fig:image_w}
\end{figure}

As a baseline for comparison, we also embed the watermark in the lower right corner of the image, as shown in the first row of~\autoref{fig:image_w}. 
In this case, we employ images from CIFAR10~\cite{cifar} as watermarks. 

To differentiate the two settings, we call them {\it text watermarks} and {\it image watermarks}, respectively.

\zztitle{Extraction Details}
For text watermarks, 
we first fine-tune MPRNet~\cite{mprNet} to reconstruct low-resolution watermarks in edited images. 
Note that due to the time-consuming nature of data collection—for example, a single \igc edit takes 10 minutes on an NVIDIA A100—we limited the fine-tuning of MPRNet to a dataset of $1,000$ images for both \ipx and \igc. 
For the second stage of our extraction process, 
we use an Optical Character Recognition (OCR\footnote{https://github.com/clovaai/deep-text-recognition-benchmark})-based model. For training our extractor \ext, we utilized {additional} $5,000$ images from \ipx and $2,000$ from \igc.

For image watermarks, we initially compare the pixel distance between the watermarked section of the image and the {\it ground truth} (\ie the selected CIFAR10~\cite{cifar} image used as the watermark) for watermark recognition. Concurrently, we utilize image reconstruction technique, \ie MPRNet, to enhance the accuracy of this method. Subsequently, we also employ a binary classification model to identify images with and without watermarks, thereby determining whether an image watermark has been added.

\begin{figure}[t]
    \centering
    \hspace*{0.6cm}
    \begin{subfigure}{0.5\textwidth}
    \includegraphics[width=7cm]{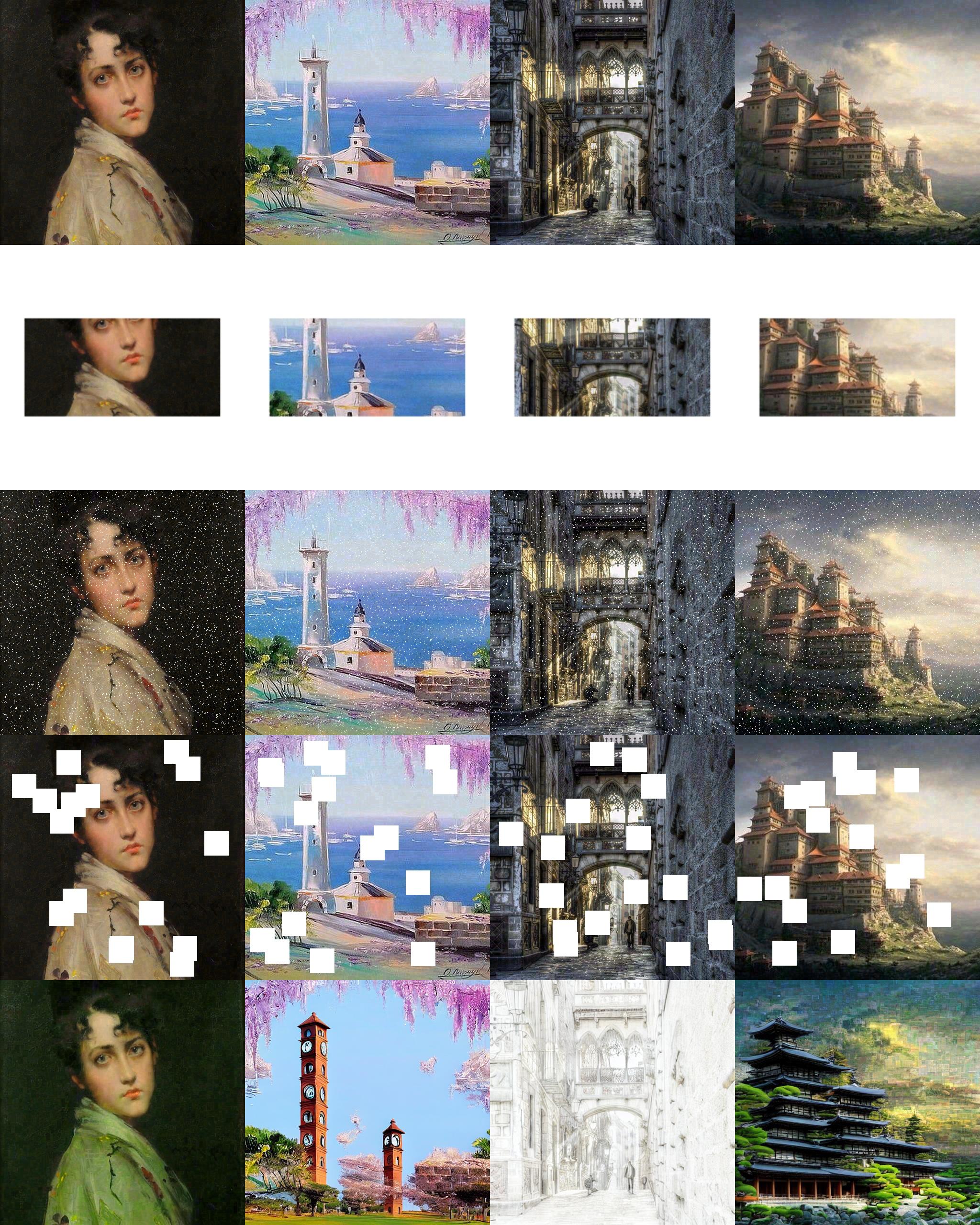}  \\
    \end{subfigure}
    \caption{Digital watermark examples: {the first row shows the results of the original images after adding a digital watermark}, rows two to four highlight traditional attacks like cropping, noise, and masking, while the last row presents post-edited images.}
    \label{fig:digital}
\end{figure}

\begin{figure*}[t]
    \centering
    \begin{subfigure}{0.32\textwidth}
        \centering
        \includegraphics[width=.98\linewidth]{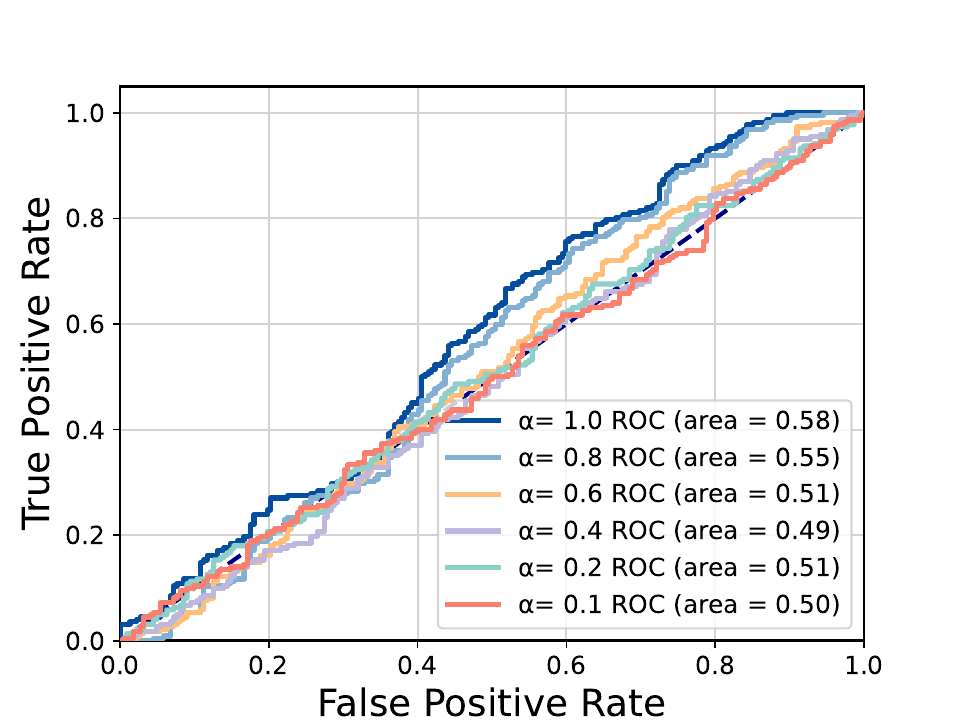} 
        \caption{Transparency (Image watermark)}
        \label{fig:image_ipx}
    \end{subfigure}
    \begin{subfigure}{0.32\textwidth}
        \centering
        \includegraphics[width=.98\linewidth]{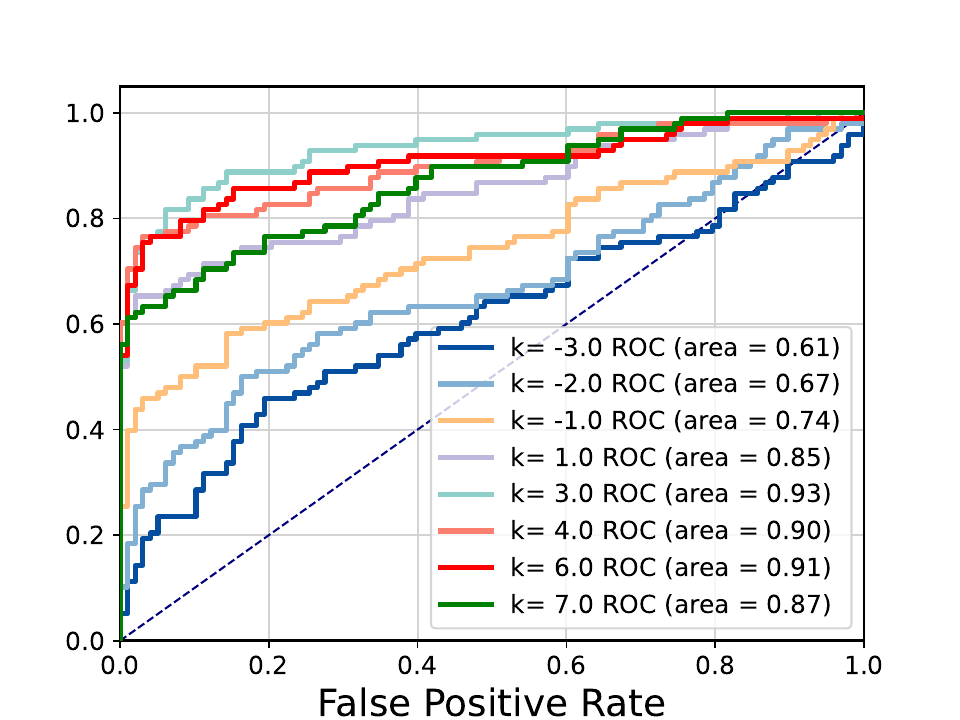}  
        \caption{Encoder Weight (Text watermark)}
        \label{fig:text_ipx_id}
    \end{subfigure}
    \begin{subfigure}{0.32\textwidth}
        \centering
        \includegraphics[width=.98\linewidth]{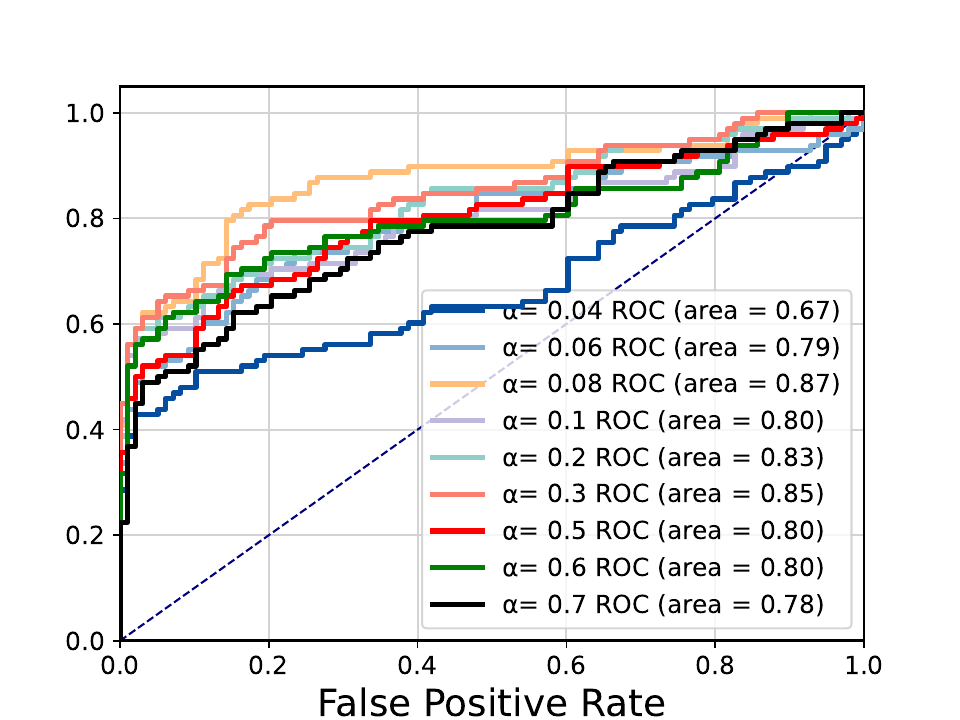} 
        \caption{Transparency (Text watermark)}
        \label{fig:text_ipx_gd}
    \end{subfigure}
    \caption{ROC results on the LAION-Aesthetics dataset and \ipx editing model. In Figure (a), we evaluate the ROC curve for naive (pixel comparison-based) extraction of image watermarking with different transparency levels $\alpha$. The clearer the image watermark is (less invisible), the better the classification results. Text watermarking (with $\mathbf{D_{letter}}$ criteria and default $\alpha=0.4,\lambda=1$), whether adjusting the encoder weights $\lambda=2^{-k}$ in (b) or the transparency of the watermark in (c), is stronger.}
    \label{fig:roc_ipx} 
\end{figure*}

\begin{figure}[t]
    \centering
    \begin{subfigure}{0.33\textwidth}
        \centering
        \includegraphics[width=.98\linewidth]{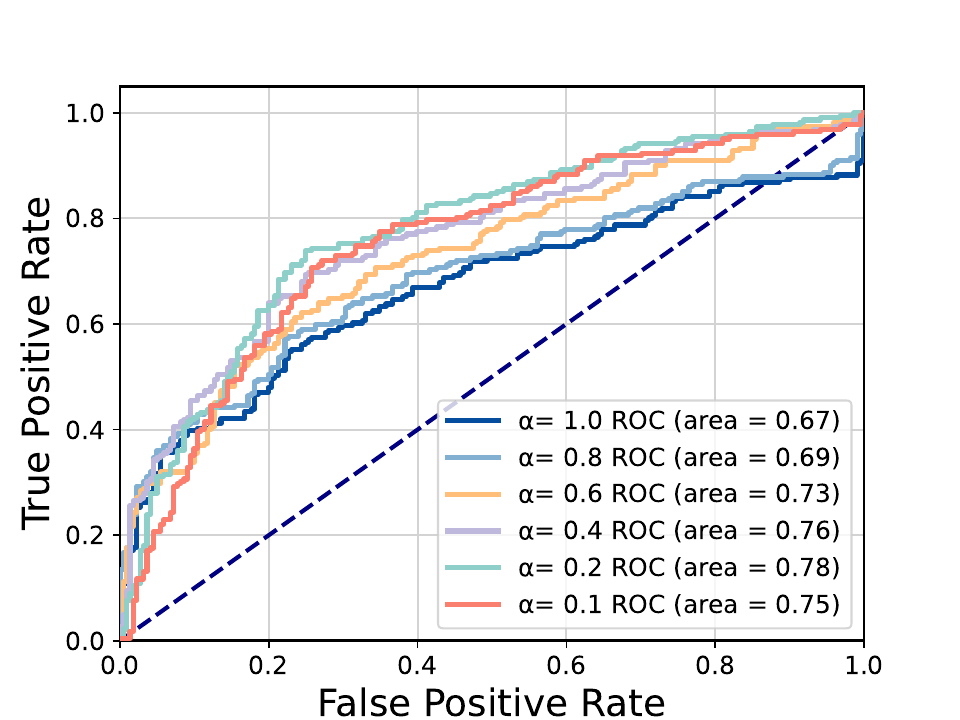} 
        \label{fig:debulr_ipx}
    \end{subfigure}
    \caption{Result of image watermarks with reconstruction using MPRNet, measured by pixel distance. 
    }
    \label{fig:debulr_roc_ipx}
\end{figure}

\begin{figure}[t]
    \centering
    \begin{subfigure}{0.33\textwidth}
        \centering        
        \includegraphics[width=.98\linewidth]{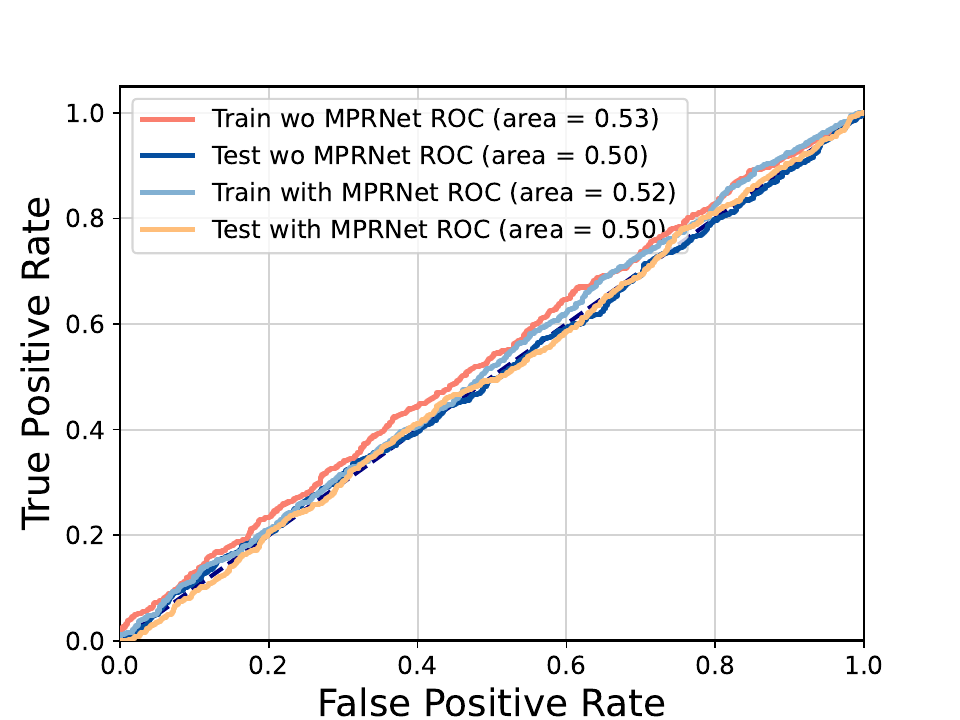} 
        \caption{without Reconstruction}
    \end{subfigure}
    \caption{{Results of a binary classification model for watermarks with and without MPRNet reconstruction.}}
    \label{fig:binary_Classify}
\end{figure}

\begin{figure}[t]
    \centering
    \begin{subfigure}{0.33\textwidth}
        \centering
        \includegraphics[width=.98\linewidth]{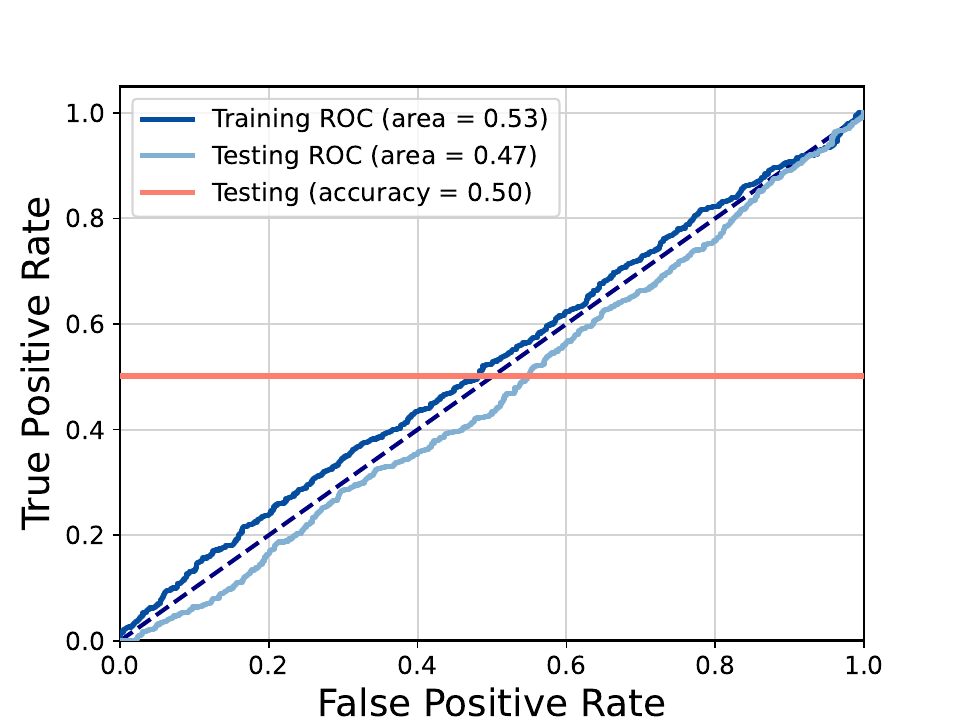} 
    \end{subfigure}
    \caption{Invisibility test. The binary classification model fails to extract sufficient watermark information from images with invisible watermarks, making it unable to differentiate them from images without any watermark.}
    \label{fig:invi_roc}
\end{figure}

\begin{figure*}[t]
    \centering
    \begin{subfigure}{0.24\textwidth}
        \centering
        \includegraphics[width=.95\linewidth]{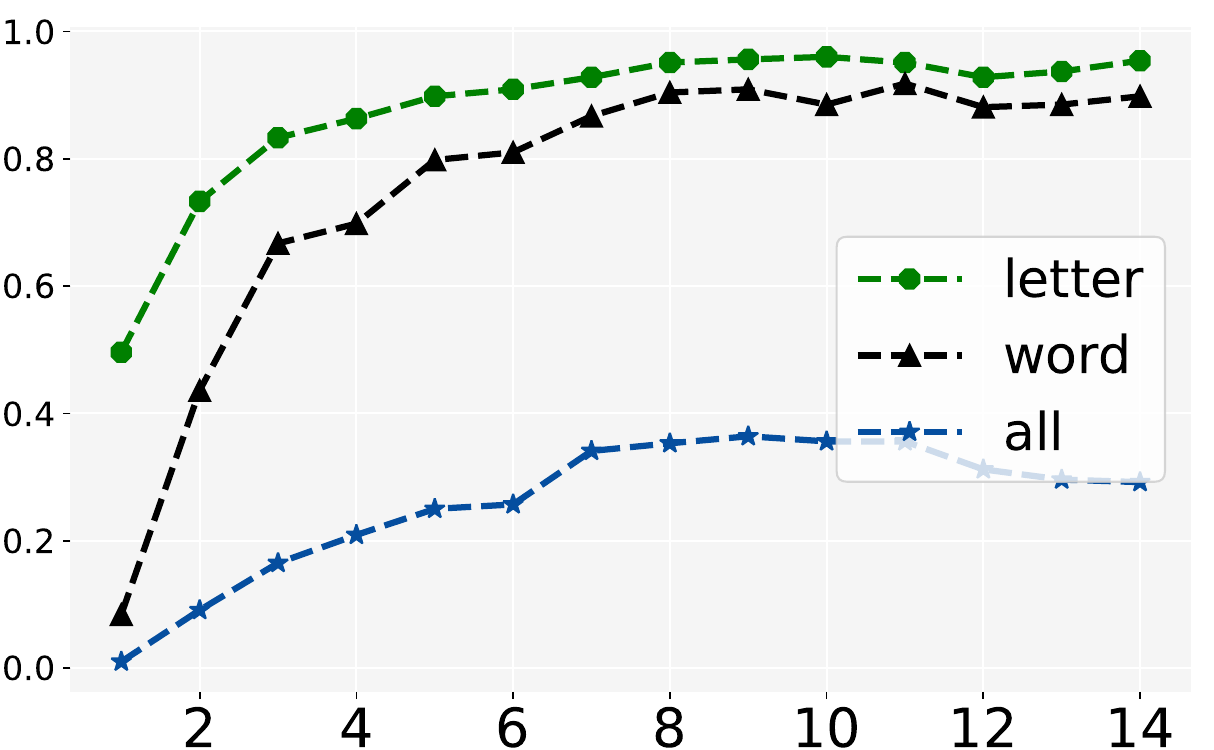}  
        \caption{Watermark Transparency}
        \label{fig:ipx_extract_a}
    \end{subfigure}
    \begin{subfigure}{0.24\textwidth}
        \centering
        \includegraphics[width=.95\linewidth]{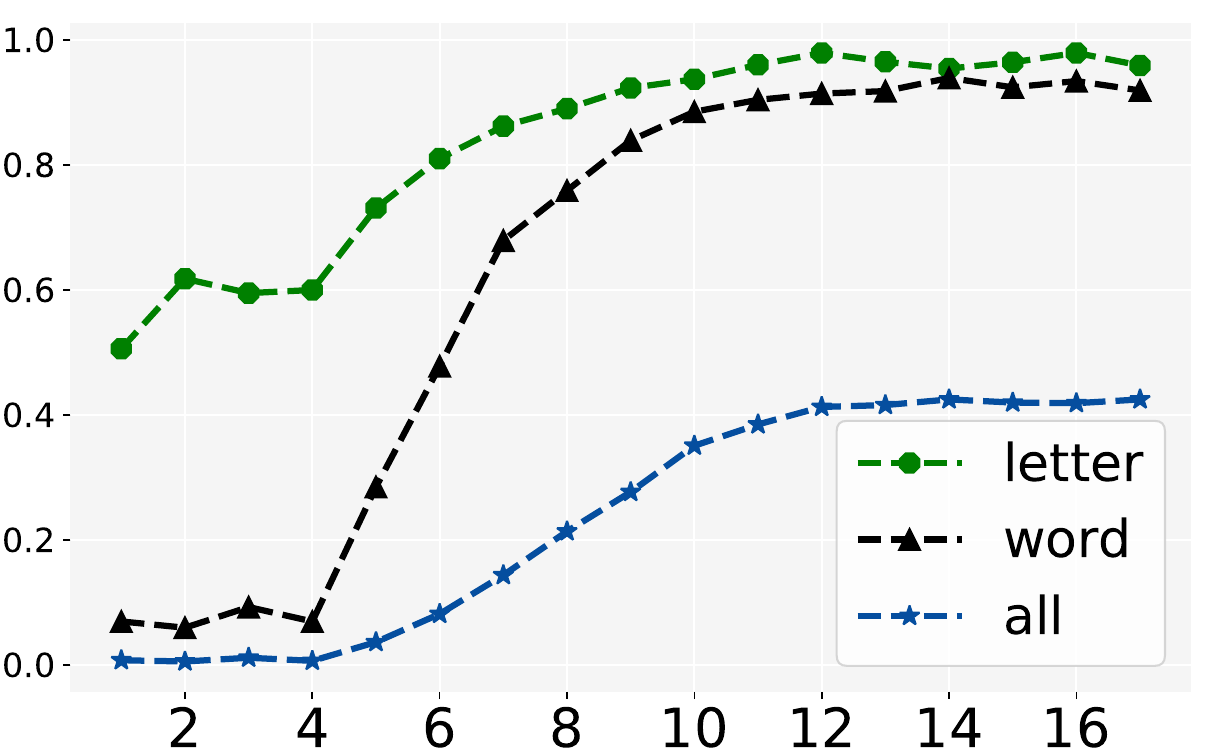}  
        \caption{Encoder Weight}
        \label{fig:ipx_extract_b}
    \end{subfigure}
    \begin{subfigure}{0.24\textwidth}
        \centering
        \includegraphics[width=.95\linewidth]{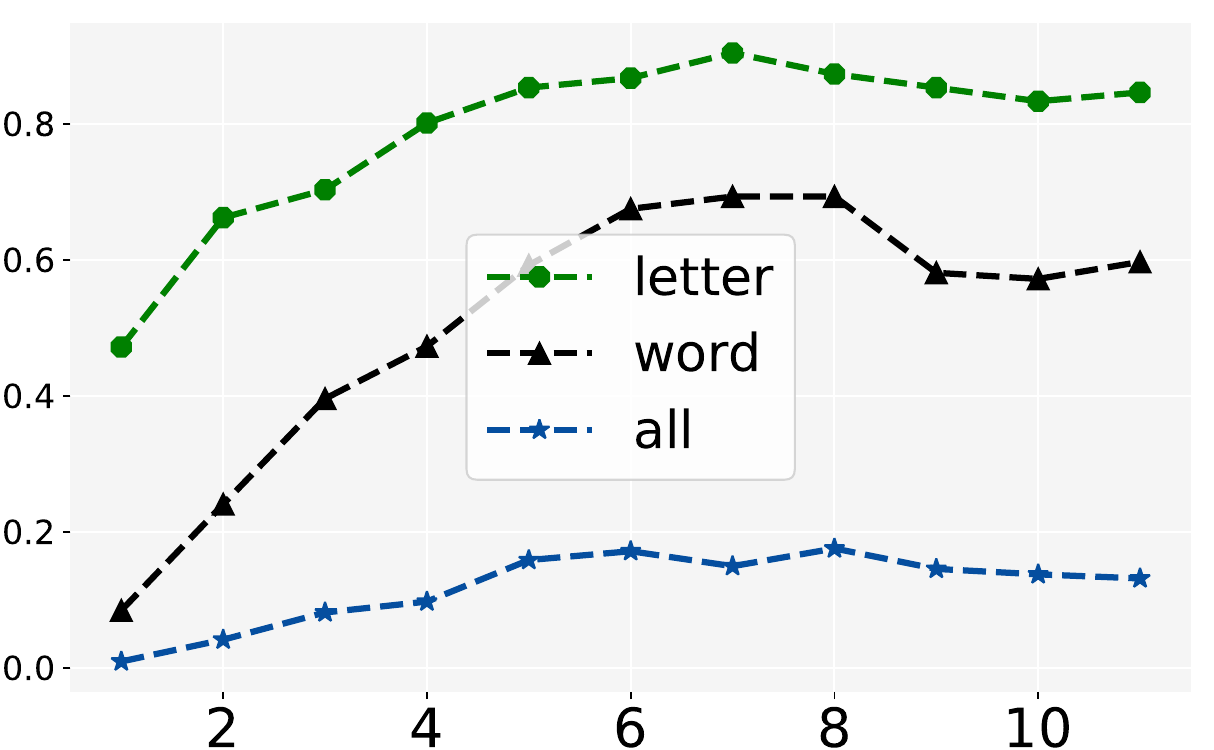}  
        \caption{Watermark Transparency}
        \label{fig:ipx_extract_c}
    \end{subfigure}
    \begin{subfigure}{0.24\textwidth}
        \centering
        \includegraphics[width=.95\linewidth]{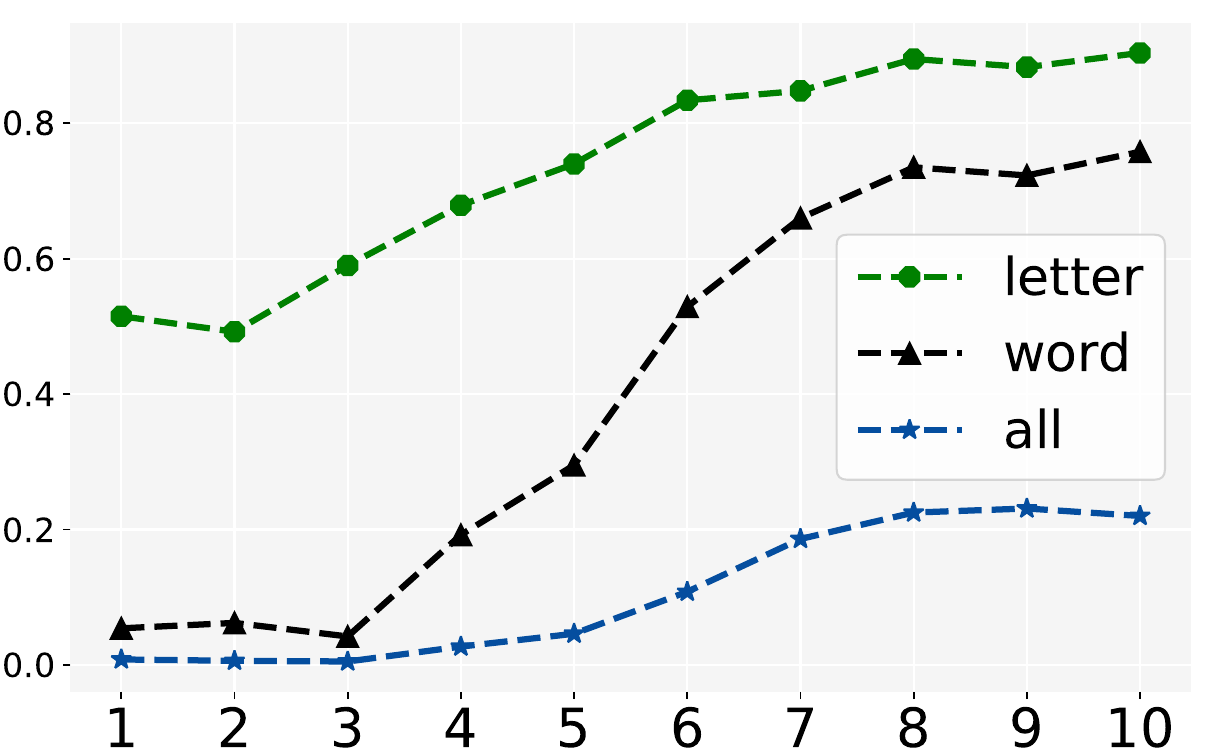}  
        \caption{Encoder Weight}
        \label{fig:ipx_extract_d}
    \end{subfigure}\\

    \begin{subfigure}{0.24\textwidth}
        \centering
        \includegraphics[width=.95\linewidth]{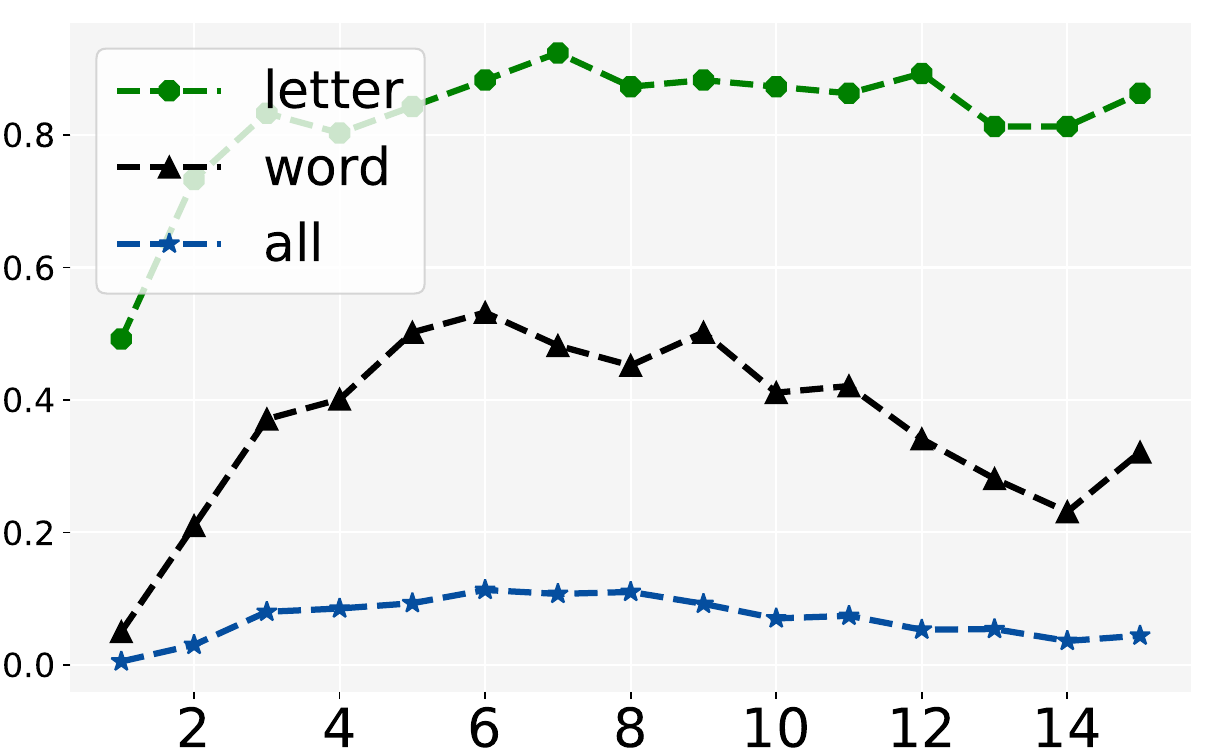}  
        \caption{Watermark Transparency} 
        \label{fig:igc_extract_a}
    \end{subfigure}
    \begin{subfigure}{0.24\textwidth}
        \centering
        \includegraphics[width=.95\linewidth]{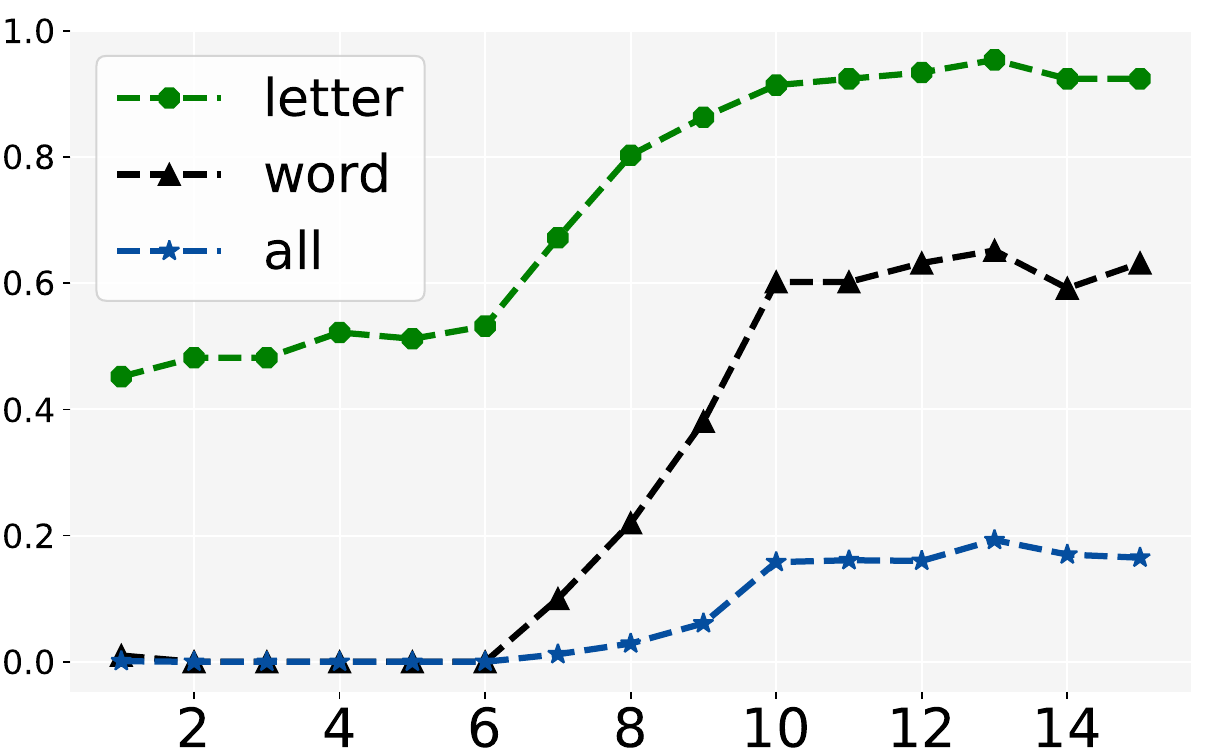}  
        \caption{Encoder Weight}
        \label{fig:igc_extract_b}
    \end{subfigure}
    \begin{subfigure}{0.24\textwidth}
        \centering
        \includegraphics[width=.95\linewidth]{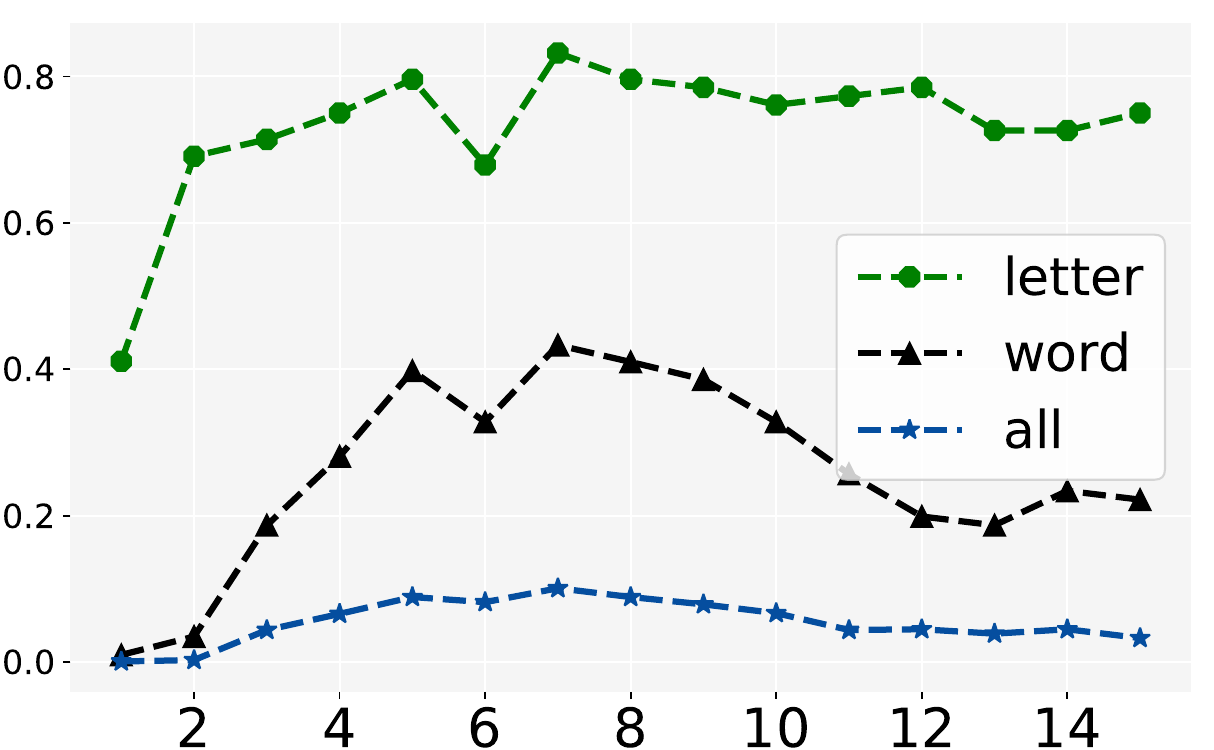}  
        \caption{Watermark Transparency}
        \label{fig:igc_extract_c}
    \end{subfigure}
    \begin{subfigure}{0.24\textwidth}
        \centering
        \includegraphics[width=.95\linewidth]{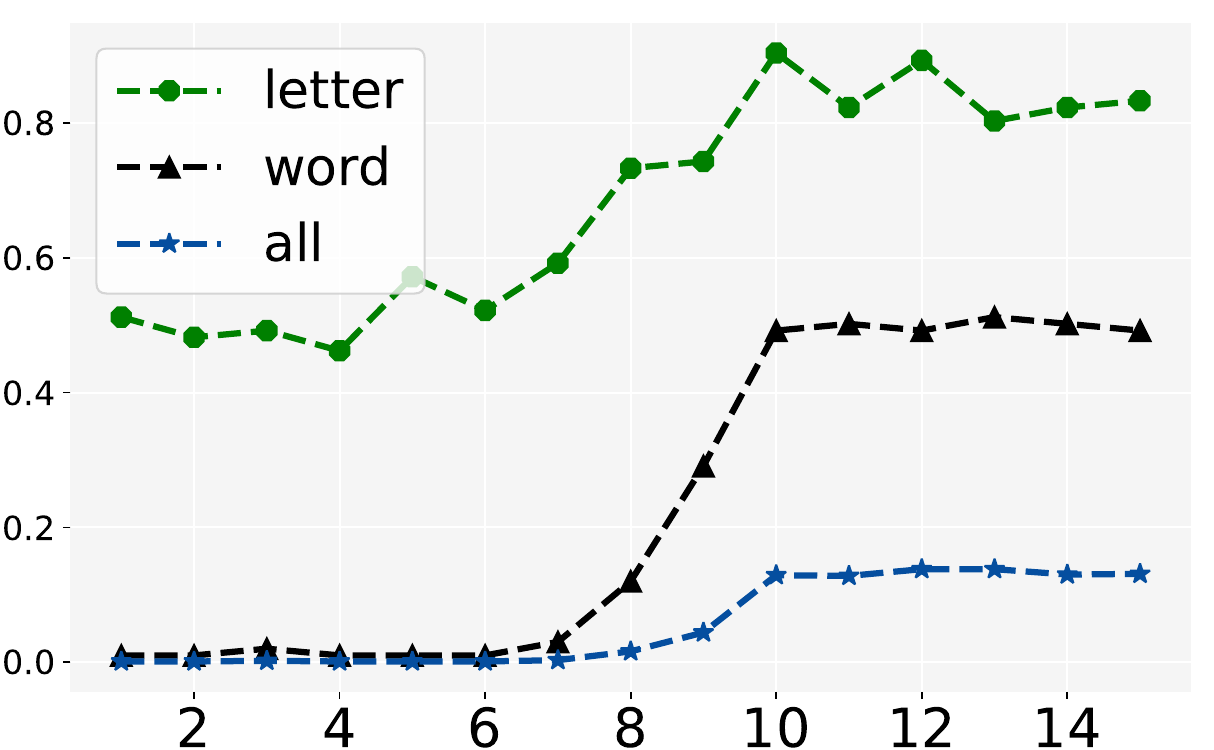}  
        \caption{Encoder Weight}
        \label{fig:igc_extract_d}
    \end{subfigure}\\
    \caption{Watermark identification accuracy varies with the increase in watermark transparency and encoder weight on the LAION-Aesthetics (above) and TEdBench (below) datasets. Subfigures (a), (b), (e) and (f) display the outcomes when edited using the \ipx model, while (c), (d), (g) and (h) showcase results from editing via the \igc model.}
    \label{fig:igc_extract}
\end{figure*}

\section{Evaluation Results}
\label{sec:eval}


\subsection{Comparison with Existing Solutions}
\label{subsec:compare}
Compared to prior watermarking methods like digital watermarks~\cite{digital_water2002,digital_water2003} and AutoEncoder-based watermarks~\cite{autoencoder_water,rivaGAN}, our approach offers notable benefits. Traditional methods yield a $0\%$ retention in editing tasks. Given the nature of the diffusion process, especially in diffusion models (DMs) tailored for editing, watermarks from these methods face extensive distortion post-editing, rendering them unextractable.

\zztitle{Digital Watermark}
We employ a popular tool\footnote{https://github.com/guofei9987/blind\_watermark} based on frequency domain techniques for inserting digital watermarks. Through evaluation, we find that the digital watermark demonstrates excellent robustness against various common image transformations such as rotation, random cropping, image masking, cutting, resizing, noise introduction, and brightness adjustment. It successfully recovers the four-letter text watermark with $100\%$ accuracy. However, when we subjected $100$ images from LAION-Aesthetics containing the digital watermark to edits using either the \ipx or \igc model, the accuracy for extracting the corresponding watermark dropped to zero.

Examples of digital watermarks are depicted in~\autoref{fig:digital}, showcasing the robustness and limitations of the digital watermark. The first row exhibits original images with the applied watermark (text ``Robust InvisiWater''). Subsequent rows (two to four) demonstrate the watermark's resilience against traditional attacks like cropping, random noise, and masking, with successful extraction still possible. However, the edited images in the final row illustrate the watermark's extraction failure. {We can also observe that the impact of traditional image transformations on images is entirely different from that introduced by edit models.} Furthermore, our analysis revealed that digital watermarks exhibit less resistance to brightness and rotation alterations than expected. A mere $15$-degree rotation or a $1.2$-fold increase in brightness rendered the watermark extraction unfeasible.

\zztitle{AutoEncoder-based Watermark} Given the popularity of utilizing AutoEncoders for watermarking tasks~\cite{autoencoder_water,rivaGAN}, particularly in recent work that employs AutoEncoder to add watermarks to diffusion models~\cite{recipe}, we also evaluate a paired Encoder-Decoder\footnote{https://github.com/yunqing-me/WatermarkDM} scheme in our study, \ie \enc and \dec. For this evaluation, we selected $10,000$ images from the LAION-Aesthetics dataset and generated $5,000$ three-letter random text watermarks for training the AutoEncoder.


We evaluate both \igc and \ipx models on $100$ images, the decoder \dec extracts watermarks with $100\%$ accuracy from unedited images. However, this accuracy plummets to $0\%$ post-editing. To enhance \dec's resilience against such manipulations, we employ adversarial training techniques aimed at improving \dec's ability to extract watermarks from perturbed images. However, the watermarks added to the images by the \enc were not retained following the editing process, resulting in $0\%$ extraction accuracy. In addition, we attempt to {\it fine-tune} the \dec that had undergone adversarial training, using images that had been edited by both \igc and \ipx models. A total of $1000$ images were used for this purpose, with $500$ images from each model. Despite the fine-tuning, there was no observable improvement in extraction accuracy, remaining $0\%$. 

\begin{table}[t]
    \centering
    \caption{Time (seconds) required by various watermarking methods to add watermarks to individual images. And pre-training or auxiliary needs for each method.}
    \label{tab:timec}
    \begin{tabular}{c|c|c|c}
        \toprule
        Watermark Tool & Digital & AutoEncoder & Ours \\
        \midrule
        Time Cost (s) & 0.85 & 0.42 & 95.03 \\
         Training-Free &  \Checkmark  & \XSolidBrush  &  \Checkmark \\
        \bottomrule
    \end{tabular}
\end{table}

\zztitle{Running Time Comparison}
In our evaluation, we assess the time overhead for watermarking an individual image using different methods, as detailed in~\autoref{tab:timec}. Notably, the AutoEncoder algorithm shows the least time overhead for watermarking. However, this does not include its substantial training time, where we spent approximately 7 hours training it on a dataset of 10,000 images using two NVIDIA A100 GPUs. In contrast, our watermarking method, though more time-consuming due to a 400-step optimization procedure, requires no additional training, data collection and pre-processing. This allows our method to be directly applied to any image without the need for preliminary steps.


\subsection{Results of Image Watermarks}\label{subsec:strength}


We opt to use images from CIFAR10~\cite{cifar} as watermarks primarily due to their widespread use in visual tasks. Extending this rationale, CIFAR10's diverse and representative image set offers a robust basis for watermarking, ensuring that our technique can be generalized across various image types and applications. In emulating traditional backdoor attack~\cite{gu2017badnets}, we strategically place the watermark in the lower right corner, a choice driven by the need for subtlety and reduced visibility. This positioning not only makes the watermark less conspicuous but also leverages the typical viewing patterns of images, where the lower right area often receives less immediate attention.



\zztitle{Extraction through Direct Pixel-wise Comparison}
{Extraction results are illustrated in~\autoref{fig:image_ipx}. Upon comparison using ROC curves and AUC values, we observe that image watermarks significantly underperform text watermarks in ~\autoref{fig:text_ipx_id} and ~\autoref{fig:text_ipx_gd} (to be discussed in \autoref{subsec:strength}). 
For instance, when the transparency of the image watermark is set to $1$, the AUC remains at a modest $0.58$, which is substantially lower than the $0.85$ AUC achieved by text watermarks with a transparency setting of $0.3$. 
A plausible explanation for this discrepancy might be the greater challenge inherent in optimizing image watermarks.
%

\zztitle{Extraction with Reconstruction}
To enhance the accuracy of watermark detection through pixel measurement, we explored using MPRNet for image watermark reconstruction. For this, we extracted the watermark portions from $1,000$ watermarked images that had undergone editing, with each portion sized $32\times32\times3$. Additionally, we used their corresponding CIFAR-10 images as the targets for model training. 

As shown in~\autoref{fig:debulr_roc_ipx}, reconstruction helps recognition.  {\it Note that there is a caveat} that we take the larger pixel-wise distance, which is supposed to be a negative case (not watermarked), as the positive (watermarked) case: On average, before reconstruction, the average distance between watermarked parts and the original CIFAR images is $147.9$. However, after reconstruction, this distance increases to $162.2$. Meanwhile, the average distance between non-watermarked parts and the original CIFAR images is $152.6$.
As a result, while the results shown in~\autoref{fig:debulr_roc_ipx} are promising, we cannot rely on them to detect watermarks. 

We believe the issue comes from the characteristics of MPRNet in image post-processing. Since the pixel measurement method is highly {\it sensitive} to noise, the distance between the watermarked image and the watermark paradoxically increased after applying MPRNet.
As shown in examples later in~\autoref{fig:edit_image} in~\hyperref[app:add_exp]{Appendix~\ref*{app:add_exp}}, image watermarks, due to the unique distribution of pixels in pictures, are completely obliterated post-editing, rendering them unrecognizable. However, for text watermarks, there is still evidence to be preserved after being edited.

\zztitle{Extraction through a Binary Classifier}
{To mitigate the negative impact of pixel distance, we further trained a binary classification model, under no MPRNet reconstruction, to discern whether image watermarks remain recognizable after editing. We collected $2,000$ images, each watermarked and edited, then extracted a $32\times32\times3$ section of the watermark from the edited images, labeling these as $1$. Similarly, we extracted watermark sections from the edited originals, labeling these as $0$. Moreover, the transparency $\alpha$ of watermarks on these $2,000$ images was randomly sampled between $0.1$ and $1.0$ to enhance the model's generalization capability. However, as shown in~\autoref{fig:binary_Classify}, we found that the model trained in this manner still failed to successfully identify image watermarks, as evidenced by an overall recognition accuracy of $0.5$, equivalent to random guessing. Subsequently, we attempted to use the MPRNet reconstruction technique, only to find that it still failed to identify the image watermark, as reflected by a model accuracy of around 0.5.



\subsection{Results of Text Watermarks}\label{subsec:strength}

\zztitle{Invisibility}
\label{subsec:invisibility}
To assert the invisibility of our watermark, we attempted to discern whether images had been embedded with an invisible watermark using a trained binary classification model. Specifically, we trained the model using $1,000$ images from LAION-Aesthetics that had been watermarked (with randomly selected watermark transparency level $\alpha$ from $[0.2, 0.4, 0.6, 0.8, 1.0]$ for generalization) and subsequently edited by \ipx (i.e., $\hat{\img}$ and $\hat{\img}_e$), along with their corresponding original images post-edit (i.e., ${\img}$ and ${\img}_e$), labelled as 1 (watermarked) and 0 (not watermarked) respectively. The test set consisted of an additional $1,000$ images. Despite various parameter tuning and model structural changes, the model could not distinguish between watermarked and non-watermarked images. This has been reflected in the test set inference results, as shown in~\autoref{fig:invi_roc}, yielding an accuracy of $0.501$ and an AUC of $0.47$.

\zztitle{Evaluation Metrics} 
%
We first establish special accuracy function $A$ for every text watermark region $\hat{\img}_k$ ($k \in K$) with function \ext as follows,
\begin{equation*}
\begin{aligned}
    A(\hat{\img}_k) =  \begin{cases} 
            1 & \text{if } \ext(\hat{\img}_k) = w \\
            0 & \text{otherwise}
            \end{cases}
\end{aligned}
\end{equation*}
Similarly, we define $A(\hat{\img}_{n,k})$ as the $k$-th watermark in the $n$-th image.
To quantify the watermark identification success in different scenarios, we define three metrics based on $A$: (1) First, we calculate the extraction accuracy rate of sub-watermarks across the entire test set as $\mathbf{D_{all}}=\frac{1}{NK} \sum_{n=1}^{N}\sum_{k=1}^{K} A(\hat{\img}_{n,k})$. (2) To measure the success rate of image ownership claims, we define metric $\mathbf{D_{word}} =\frac{1}{N} \sum_{n=1}^{N}\min \left \{  \sum_{k=1}^{K} A(\hat{\img}_{n,k}),1 \right \}$.
The implication of this metric is that among the $K$ identical independent watermarks we introduce, if even one is successfully recognized, it is deemed as proof of ownership. (3) Finally, we introduce metric $\mathbf{D_{letter}}$ to measure the identification success rate of special character watermarks. This method allows developers to use unique symbols as watermarks to stake their ownership claims. For instance, in an individual watermark containing $M$ characters, only all $M$ symbols are successfully identified within an image, and can be combined into the entire added symbol sequence, we consider the watermark recognition to be successful. 


As elaborated in~\autoref{subsec:iij}, our text watermark's strength in target images affects its visibility in edited images. The watermark's transparency and the weight of the encoder loss in our adversarial training both impact post-editing watermark extraction. Hence, we assess the impact of varying watermark transparency and encoder loss weights on extraction accuracy in the two editing models. 
Note that we ignore the false positives as we observe {low average false positive rates of $0\%$, $0\%$, and $25.5\%$ on various $\alpha$, in LAION-Aesthetics dataset, across the three metrics for the no-watermark cases.} In addition, to comprehensively assess the validity of our \ext, we calculated the ROC curves for \ipx under the $\mathbf{D_{letter}}$ mode across different values of $\lambda$ and $\alpha$. The results are displayed in ~\autoref{fig:text_ipx_id} and~\autoref{fig:text_ipx_gd}. When $\lambda=1/8$ and $\alpha=0.3$, the AUC values reach $0.93$ and $0.85$ respectively, indicating a certain degree of stability in the method.

\zztitle{Results on LAION-Aesthetics} Firstly, we evaluate the performance of our \invinj on the LAION-Aesthetics dataset. As shown in~\autoref{fig:ipx_extract_a} and~\autoref{fig:ipx_extract_c}, we observed that as the transparency of the watermark in the target image $\img'$ increases, the extraction accuracy correspondingly improves across both \ipx and \igc editing model. Moreover, it was found that beyond a certain level of transparency, specifically at around $102/255$, the extraction accuracy across \ipx and \igc begin to stabilize at $96.0\%$ and $90.4\%$, respectively. As evident from~\autoref{fig:ipx_example}, when the transparency reaches $102/255$, the watermark we injected becomes visually imperceptible in the edited images. Meanwhile, even when the transparency is as low as $15/255$, the watermark can still be successfully extracted. Similar results were observed with adjustments to the encoder weight $\lambda$. As illustrated in~\autoref{fig:ipx_extract_b} and~\autoref{fig:ipx_extract_d}, when we increase $\lambda$ to $1/2$ or lower, our $\mathbf{D_{letter}}$ success rate stabilizes at $97.9\%$ and $90.4\%$. 


\zztitle{Results on TEdBench} Similar results are observed on the TEdBench dataset with both two editing models, as shown in~\autoref{fig:igc_extract_a} and~\autoref{fig:igc_extract_c}. Once the watermark transparency for $\img'$ reaches $51/255$, watermark identification accuracy levels off. In both the \ipx and \igc models, when the watermark transparency is $51/255$, the identification accuracy reaches respective accuracies of $92.4\%$ and $83.2\%$. Additionally, as shown~\autoref{fig:igc_extract_b} and~\autoref{fig:igc_extract_d}, when $\lambda$ is below $1/2$, identification accuracy progressively becomes stable, attaining values of $91.4\%$ and $89.3\%$ respectively when $\lambda$  is equal to $1/2$. The TEdBench dataset, with $50$ images and $100$ editing instructions, is inadequate for training \ext, which was trained exclusively on LAION-Aesthetics. This contributes to the observed lower watermark identification accuracy on TEdBench. A dip in accuracy occurs, as seen in~\autoref{fig:igc_extract_c}, when watermark transparency surpasses $102/255$, likely due to \ext's training with watermarks in $\img'$ at this default transparency. Still, \ext demonstrates satisfactory generalizability, with \igc editing results in~\autoref{fig:imagic_exam}.

\begin{table}[t]
    \centering
    \caption{Improvement in watermark identification accuracy for both editing models \ipx ({Ipx}), \igc ({Igc}) and datasets LAION-Aesthetics (LA), TEdBench (TB) under conditions of low watermark clarity when introducing reconstruction. Specifically, these results are obtained when transparency $\alpha$ is set to $1/255$ and  $\lambda$ is $5e10$. Both of them control watermark strength (Stren). In the table, ``w/o Rec'' denotes results without reconstruction, while ``w/ Rec'' indicates results with reconstruction. $\Delta$ indicates improvement of the reconstruction.} 
    \label{tab:reconstruct}
    \begin{tabular}{c|cc|cccc}
    \toprule
    Edit                    &   {Dataset}                        & Stren       &  \textbf{w/o Rec}    &  \textbf{w/ Rec}  & $\Delta$     \\ \midrule
    \multirow{4}{*}{{Ipx}  }  & \multirow{2}{*}{{LA}}        & $\alpha$    &  $49.6\%$  &  $54.3\%$  & $\mathbf{9.5\%}$    \\ 
                                   &                             &$\lambda$  &  $50.6\%$  &  $55.7\%$  & $\textbf{10.1\%}$   \\
                            & \multirow{2}{*}{{TB}}              &$\alpha$     &  $49.2\%$  &  $49.2\%$  & $0\%$  \\ 
                            &                                    &$\lambda$  & $45.2\%$   &  $45.2\%$  & $0\%$  \\  \hline  
    \multirow{4}{*}{{Igc} }  & \multirow{2}{*}{{LA}}        &$\alpha$     &  $47.2\%$  &  $50.7\%$  & $\textbf{7.4\%}$    \\ 
                            &                                    &$\lambda$  &  $51.5\%$  &  $57.0\%$  & $\textbf{10.7\%}$   \\
                            & \multirow{2}{*}{{TB}}              &$\alpha$     &  $41.1\%$  &  $42.1\%$  & $\textbf{2.4\%}$  \\ 
                            &                                    & $\lambda$ & $51.2\%$   &  $52.2\%$  & $\textbf{2.0\%}$  \\  \bottomrule 
    \end{tabular}
\end{table}

\begin{figure}[t]
    \centering
    \includegraphics[width=.45\linewidth]{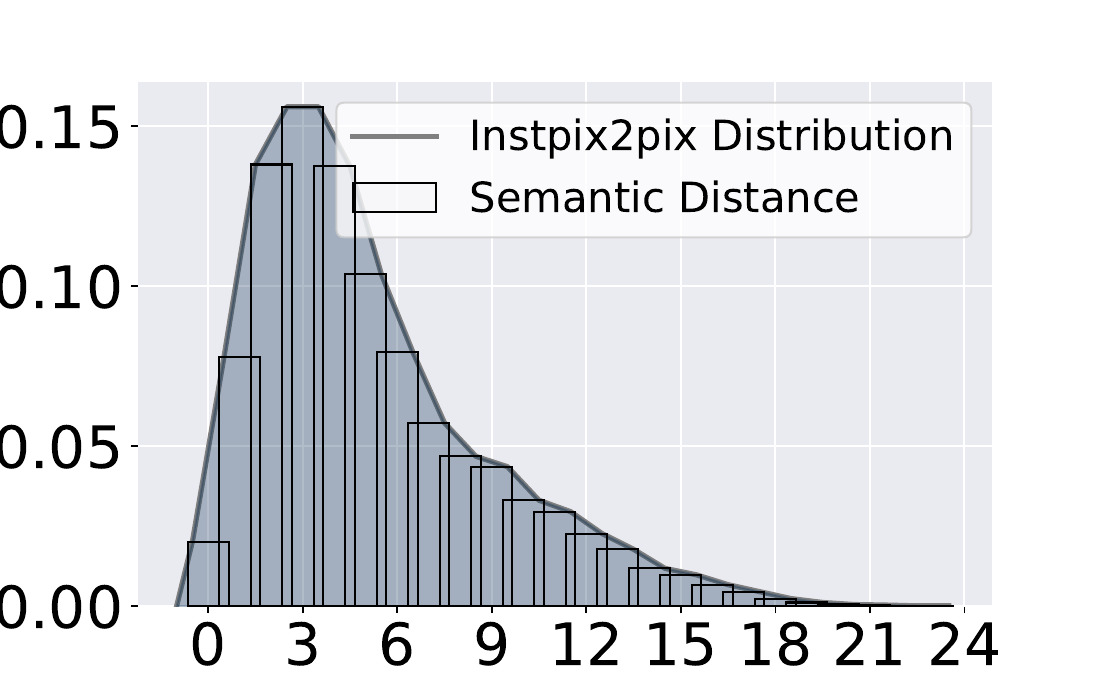}  
    \includegraphics[width=.45\linewidth]{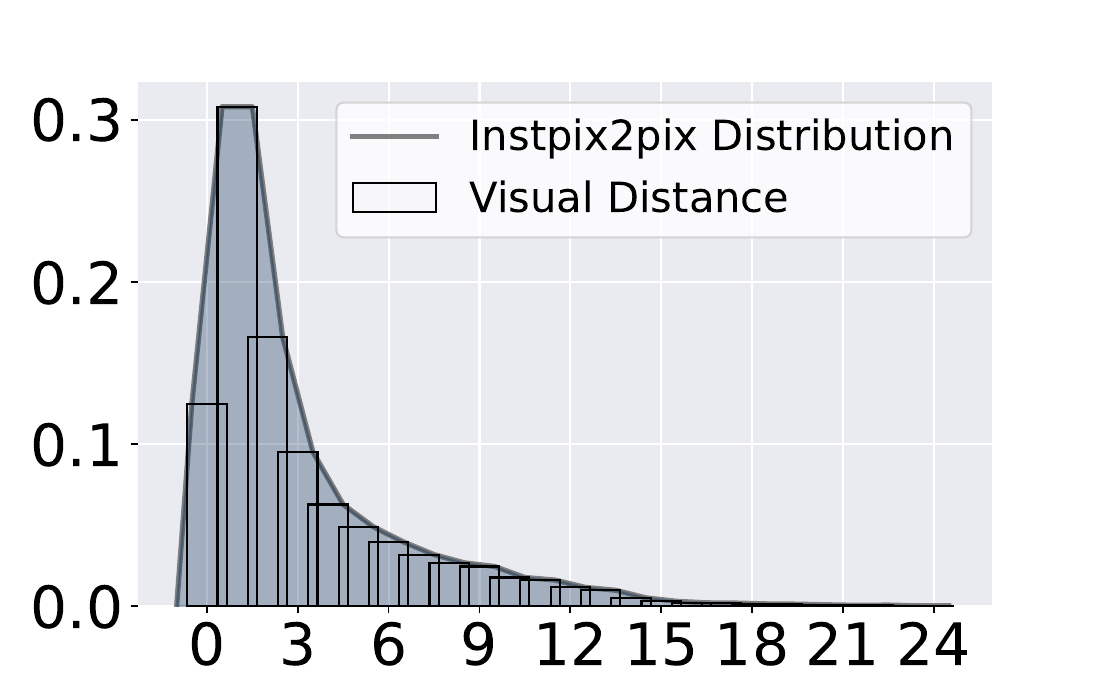}  
    \includegraphics[width=.45\linewidth]{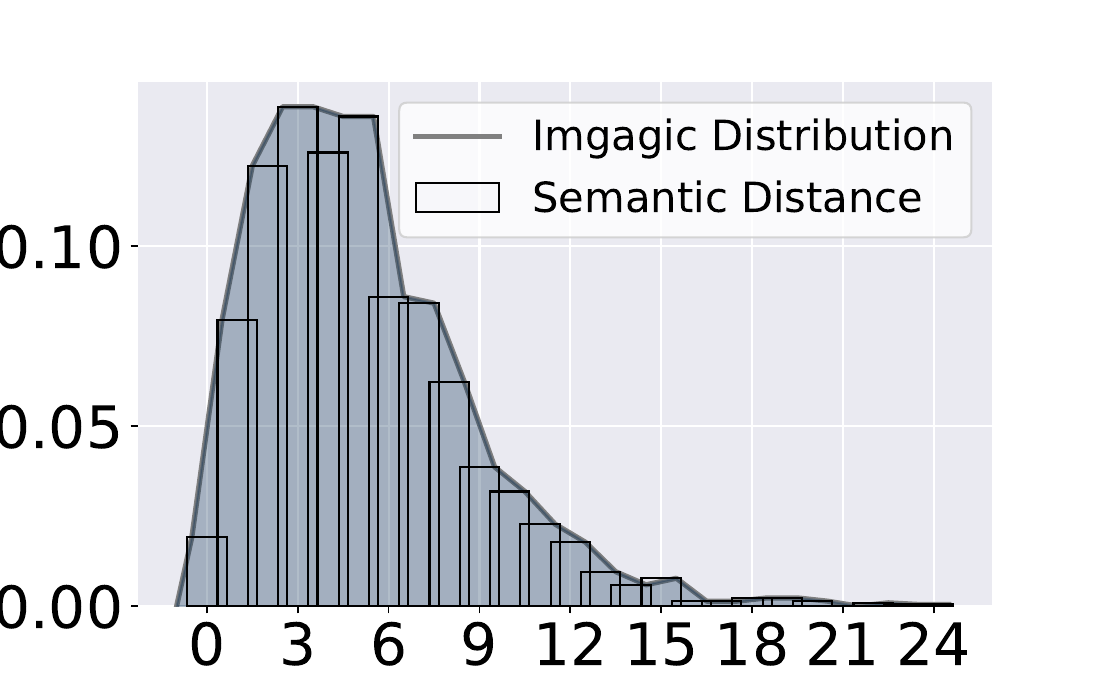}  
    \includegraphics[width=.45\linewidth]{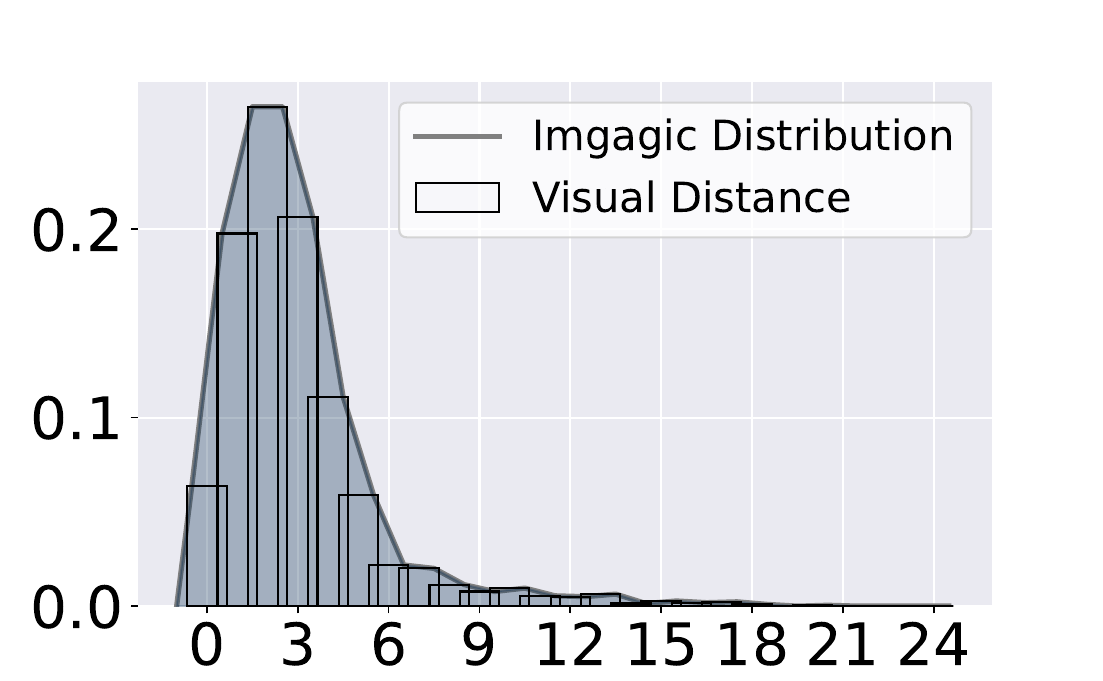} 
    \caption{Distribution of semantic and visual changes before and after editing with the \ipx and \igc models. The first row shows the distribution for the \ipx model with semantic and vision distance (very close), while the second row shows those for the \igc model. 
    }
    \label{fig:edit_dist}
\end{figure}
\zztitle{Impact of Encoder Weight} We evaluate the impact of changes in the weight of encoder loss on the invisibility of watermarks after being edited. As $\lambda$ approaches $1/2$, the watermark extraction accuracy stabilizes, as shown in~\autoref{fig:igc_extract}. However, when $\lambda$ reaches $1/128$, we can faintly discern a subtle watermark after editing. 
As examples shown in~\autoref{fig:ipx_encode} in~\hyperref[app:add_exp]{Appendix~\ref*{app:add_exp}}, users can adjust the watermark strength parameter in our optimization function to determine the visibility of the watermark after the image is edited. When the watermark is conspicuous, it serves as a warning against malicious edits. When the watermark is invisible, it assists in claiming the copyright of the edited image.
\subsection{Benefit of Reconstruction}
\label{subsec:reconstruct}
We evaluate the efficacy of the reconstruction component in our method. We observe that under conditions where watermark clarity is extremely low, the reconstruction can be beneficial. For instance, on the LAION-Aesthetics dataset using the \ipx, when watermark transparency is set to $1/255$ or $\lambda$ is set to $5e4$, there is an increase in identification accuracy by $9.5\%$ and $10.1\%$, respectively. 
For a more comprehensive view of results across two edit models and two datasets of low clarity watermark shown in~\autoref{tab:reconstruct}. When the extraction accuracy of text watermarks is already high, the improvement brought about by reconstruction techniques is not significant. Furthermore, watermark reconstruction technology is more effective for text watermarks than for image watermarks, reflecting certain limitations of the reconstruction model MPRNet~\cite{mprNet}.

\begin{figure*}[t]
    \centering
    \begin{subfigure}{0.48\textwidth}
        \centering
        \includegraphics[width=.95\linewidth]{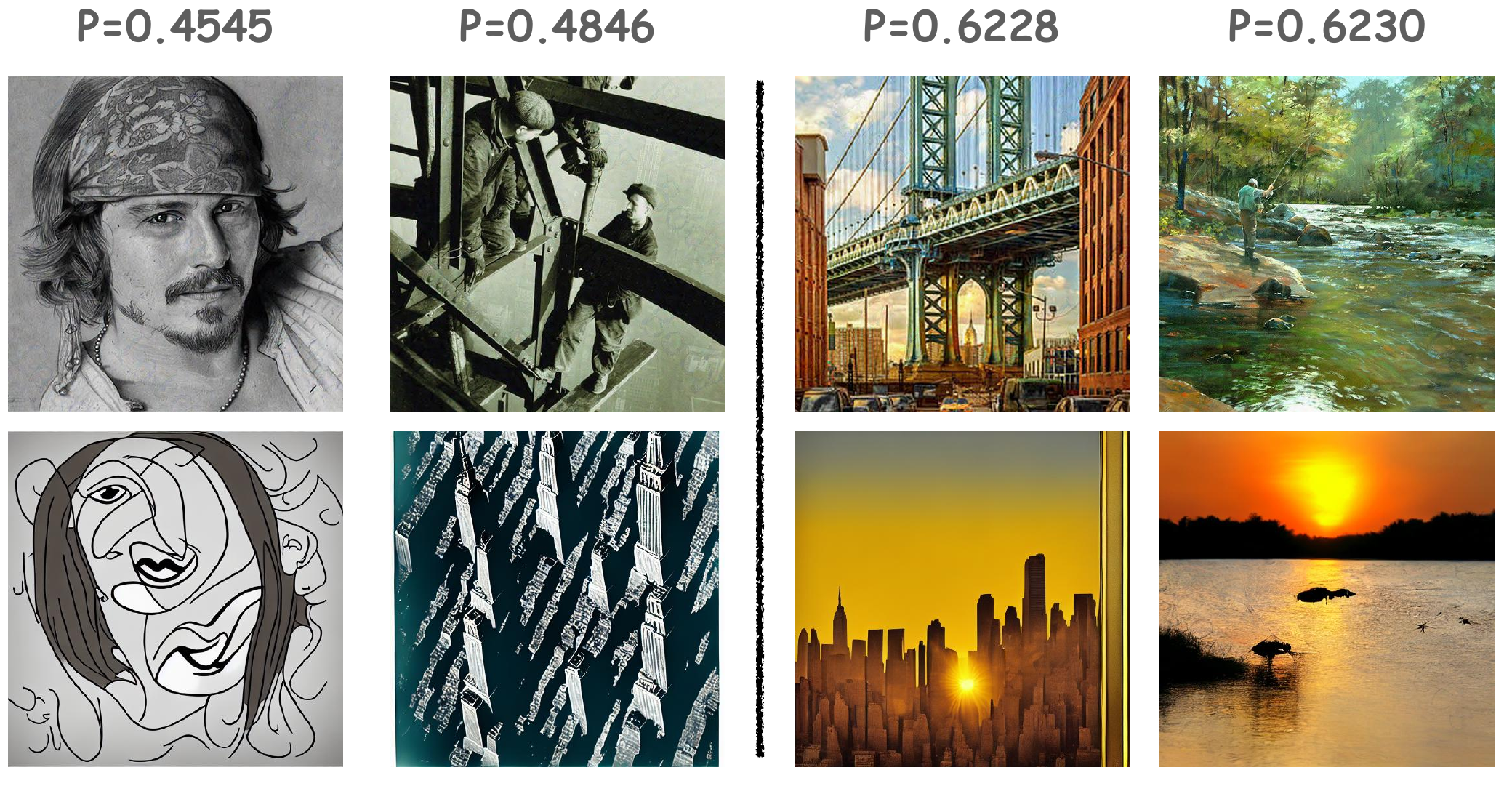}
        \caption{Examples edited by \ipx}
        \label{fig:imp_ws_mentr_a}
    \end{subfigure}
    \begin{subfigure}{0.48\textwidth}
        \centering
        \includegraphics[width=.95\linewidth]{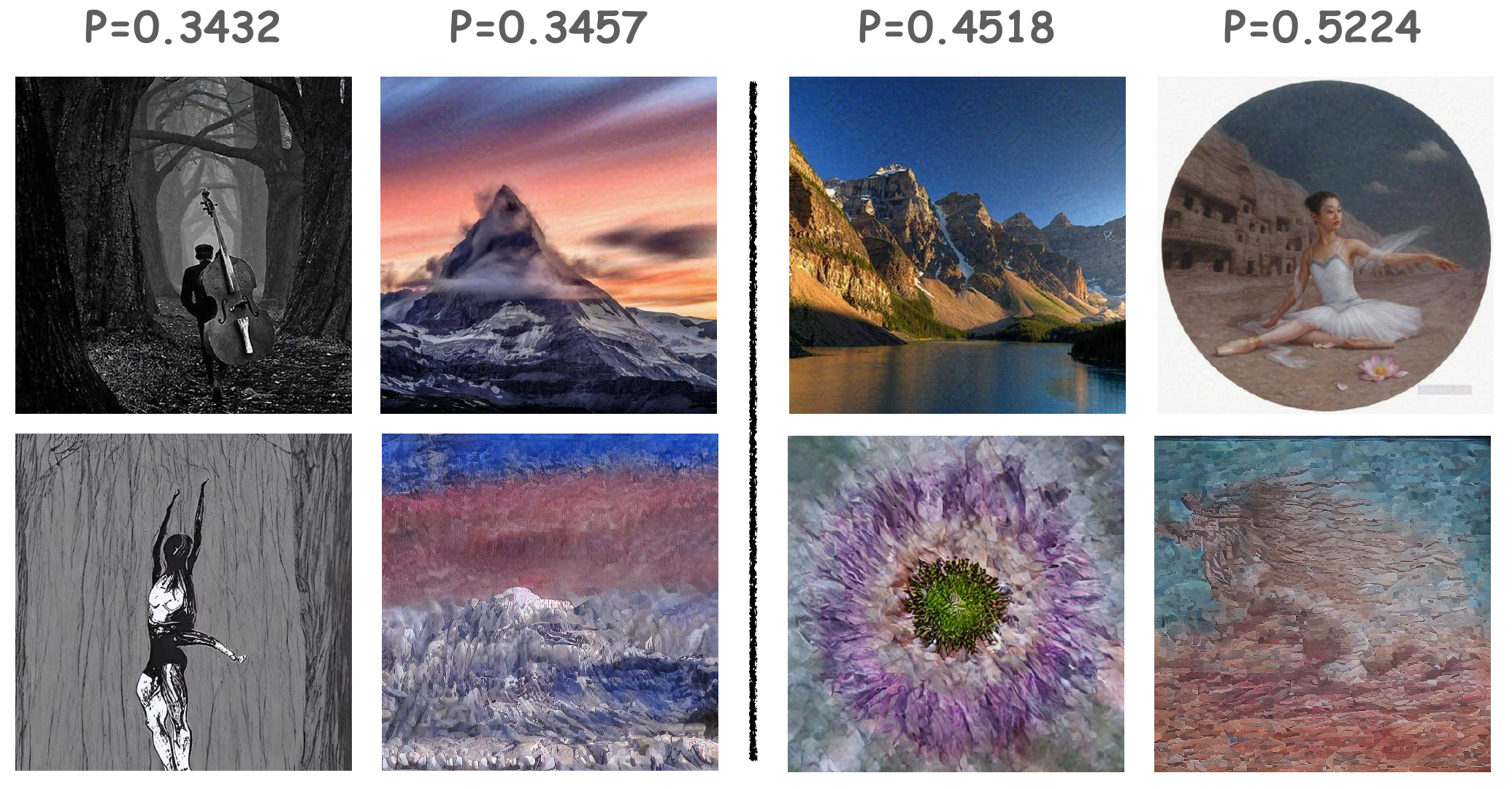}  
        \caption{Examples edited by \igc}
        \label{fig:imp_ws_mentr_b}
    \end{subfigure}
    \caption{Examples with notable semantic and vision distances before (first row) and after (second row) editing by the \ipx and \igc edit models. In each subfigure, the left panel represents semantic distance while the right shows vision distance. P indicates the corresponding distance values.}
    \label{fig:dis_example}
\end{figure*}


\subsection{Watermark Protection Boundary}
\label{subsec:boundary}
Considering that editing models can modify an image's distribution and semantics, preserving ownership amid significant changes to the image's content or style becomes crucial. Thus, we assess the degree of alteration in the image's semantic or visual content after editing with either the \igc or \ipx models. As the distance in both semantic and visual dimensions increases, the watermark becomes increasingly difficult to extract due to the loss of original image information. {\autoref{fig:edit_dist} displays the distribution of changes in images before and after editing. The two images on the left represent the semantic distance based on CLIP~\cite{clip}, while the two images on the right show the visual distance calculated using VGG~\cite{vgg}.} {Specifically, akin to existing works~\cite{clip,lpips,schroff2015facenet}, when calculating distances using CLIP and VGG, we first project the images onto embeddings and then compute the $\ell_2$ distance between these embeddings.}

As observed from~\autoref{tab:dist_acc}, when we incorporate the top $5\%$ of images with the largest changes in semantic and vision distance into our analysis, there is a varying degree of decline in the accuracy of different watermark extraction strategies. Notably, the \ipx model tends to have a more significant impact than the \igc. This is attributed to the fact that \ipx induces a larger shift in the image distribution, as also indicated by the distance threshold (Thred), \ie $46.7\%$ and $47.5\%$ higher in semantic and vision distance than \igc. The overall accuracy of text watermark extraction, as it varies with the edit distance, is illustrated by~\autoref{fig:dist_acc_trend} in~\hyperref[app:add_exp]{Appendix~\ref*{app:add_exp}}.

\begin{table}[t]
    \centering           
    \caption{Watermark extraction accuracy for the top $5\%$ based on the Semantic (Sem) and Vision (Vis) distance (Dist) before and after editing with \ipx ({Ipx}) and \igc ({Igc}) respectively, along with the specific distance threshold (Thred).} 
    \label{tab:dist_acc}
    \begin{tabular}{c|cc|ccc}
    \toprule
    {Edit}       &   {Dist}           &  {{Thred}}            &  $\mathbf{D_{letter}}$$\downarrow$    &  $\mathbf{D_{word}}$$\downarrow$   & $\mathbf{D_{all}}$$\downarrow$    \\ \midrule
    \multirow{2}{*}{{Ipx} }  &   {Sem}                & $0.44$                              &  $0.21\%$                &  $0.51\%$              &  $2.73\%$    \\ 
                            &   {Vis}                  & $0.62$                              &  $0.23\%$                &  $0.56\%$              &  $3.27\%$  \\ 
                            \Xcline{2-6}{0.5pt}
    \multirow{2}{*}{{Igc}}  &   {Sem}                & $0.30$                              &  $0.06\%$                &  $0.11\%$              &  $0.07\%$    \\ 
                            &   {Vis}                  & $0.42$                              &  $0.11\%$                &  $0.32\%$             &   $0.00\%$   \\  \bottomrule
    \end{tabular}
\end{table}

As illustrated in~\autoref{fig:dis_example}, a detailed comparison of the visual changes between the original and edited images, based on the two different models, reveals a notable observation. When the distance between the original and edited versions is substantial, the alterations in the images are quite significant. This leads to the edited images retaining minimal resemblance to their original counterparts. Consequently, such extensive modifications raise a question about the necessity of further discussions regarding copyright protection. The stark contrast in visual appearance suggests that the edited images might be viewed as distinct enough to mitigate concerns over copyright infringement.




\section{Conclusion}
\label{sec:conclusion}
\vspace{-0.09cm}
In this paper, we are the first to address the issue of image copyright and malicious alterations within the context of diffusion model-based editing frameworks. We have formalized this issue and presented it within a Game framework. Subsequently, we conduct an exhaustive study of prior watermarking techniques, such as digital watermarks or AutoEncoder-based watermarks. We discover that traditional watermarking approaches, even when enhanced, are inadequate to handle the challenges introduced by the editing models of the diffusion era. Building on this, we propose our invisible watermarking method \invinj (Robust Invisible Watermarking) based on visual-semantic information. Compared to traditional watermarking techniques with $0\%$ retention, our method maintains a $96.0\%$ watermark extraction accuracy post-editing. This ensures that our watermark effectively addresses the image copyright and deceptive editing concerns in the editing model framework.


\bibliographystyle{plain}
\bibliography{main}

\appendices
\section{Additional Background} \label{appendix:A}
We first present Denoising Diffusion Probabilistic Models (DDPMs)~\cite{ddpm}, which serve as the foundation for image editing models, and then, we dive into detailed descriptions of two state-of-the-art and open-sourced edit models we use in this paper.

\subsection{Denoising Diffusion Probabilistic Models}
\label{subsec:dms_edit}

In DDPMs, a data item $x_0$, sampling from the original distribution, undergoes a forward diffusion process to gradually approach the Gaussian distribution $N(0,1)$. The reverse diffusion starts with a noise image sampled from $x_T \sim N(0,1)$ and progressively refines it into a photorealistic image $x_0$. Each intermediate sample, $x_t$, for $t \in { 1,...,T }$, adheres to the format,
\begin{equation*}
    x_t=\sqrt{a_t} x_0+\sqrt{1-a_t}\varepsilon _t
\end{equation*}
with the hyper-parameter $a_t \in (0,1)$ of the diffusion schedule and $\varepsilon_t \sim N(0,I)$, each refinement step is predicted using the neural network model $f_{\theta}$. Therefore, $x_{t-1}$ is derived from $f_{\theta}(x_t,t)$, followed by a Gaussian perturbation as, 
\begin{equation*}
    x_{t-1}=f_{\theta}(x_t,t)+ \sigma _tz
\end{equation*}
with $z \sim N(0,I)$. The training objective of $f_{\theta}$ is optimized by variational lower bound as~\cite{ddpm,nonequilibrium},  
\begin{equation*}
    L_t=E_{t\sim[1,T],x_0,\varepsilon _t}\left [ \left \| \varepsilon _t-f_\theta (\sqrt{a_t} x_0+\sqrt{1-a_t}\varepsilon _t,t) \right \| ^2  \right ] 
\end{equation*}
DDPM can be adapted to learn conditional distributions by conditioning the denoising neural network $f_\theta$ on an auxiliary input $y$. Consequently, the network $f_\theta(x_t,t,y)$ samples from a data distribution based on $y$. Typically, $y$ is a descriptive text for the training image~\cite{dalle,ediffi,dreambooth,imagen}. However, $y$ can also represent a low-resolution image~\cite{image_condition} or a class label in ImageNet~\cite{class_condition}. For our editing tasks, $y$ encompasses both the edited image and text prompt~\cite{ipx,imagic}.

\begin{table*}[t]
    \caption{Different style of text edit instruction between \ipx ({ipx})~\cite{ipx} and \igc ({igc})~\cite{imagic}.}\label{tab:edit_style}
    \resizebox{1\textwidth}{!}{
    \small
    \centering
    \begin{tabular}{cccc}
    \toprule
                                &                                    & \textbf{Romantic Venice Oil Paintings}                 &  \textbf{A photo of Barack Obama}                       \\ \hline
    \multirow{2}{*}{Instruction Style}  & {ipx} & Add painting a helicopter                              & Turn Barack Obama into smiling with a big grin  \\
                                      & {igc}   & Romantic Venice with Helicopter Oil Paintings & A photo of Barack Obama smiling with a big grin  \\
    \bottomrule
    \end{tabular}
    }
\end{table*}

\begin{table}[htbp]
    \centering
    \caption{An automated dialogue to generate edit instructions based on GPT-4-0613 API, tailored to match the prompt styles of various editing models.}
    \label{tab:instGen}
    \begin{tabularx}{0.49\textwidth}{X}
        \toprule
        \textbf{Prompt}: Let's play a game. I'll provide a sentence A, which describes an image, for example: ``the little prince fox wallpaper".  Then, I'll give a modification instruction C for that image, such as ``make the fox roguish". The outcome will be a modified description, sentence B, which could be: ``roguish fox wallpaper". In our game, I'll supply you with sentences A and B, and your task is to deduce the corresponding modification instruction, sentence C. Require C natually like the image editing instruction.\\
        \textbf{GPT4}: {\tt Sounds like an interesting challenge, let's go for it! }\\
        \\
        \textbf{Prompt}: Now, I provide the sentence A ``A photo of a standing dog'' and sentence B ``A photo of a sitting dog''. Now answer me with the corresponding instruction sentence C. \\
        \textbf{GPT4}: {\tt Change the standing dog into a sitting position.} \\
        \\
        \textbf{Prompt}: Now, I provide the sentence A ``A photo of a tennis ball'' and sentence B ``A photo of a tomato in a blue tennis court''. Now answer me with the corresponding instruction sentence C. \\
        \textbf{GPT4}: {\tt Replace the tennis ball with a tomato in a blue tennis court.} \vspace{0.2cm}
        \\
        \multicolumn{1}{c}{......}\vspace{0.3cm}
        \\
        \textbf{Prompt}: Now, I provide sentence A ``A photo of a teddy'' and sentence B ``A photo of a teddy bear doing pushups''. Now answer me with the corresponding instruction sentence C. \\
        \textbf{GPT4}: {\tt Make the teddy bear look like it's doing pushups.} \\
        \bottomrule
    \end{tabularx}
\end{table}

\subsection{Details about Editing Models}
\label{subsec:edit_models}
\zztitle{\ipx} 
It utilizes a pre-trained Stable Diffusion (SD)~\cite{latent_space} model as the base model. SD incorporates DDPM~\cite{ddpm} for the diffusion process and CLIP~\cite{clip} for semantic alignment with textual descriptions. The image inference process involves projecting images into the {\it latent space} using an encoder \enc, and then restoring them back to images with a corresponding decoder \dec. 
During the editing process, the target image for editing, $\img$, and the editing instructions, $p$, are fed into SD. 
The edited image is then gradually inferred from random Gaussian noise $z_t$, with each step $t$ of inference targeting to achieve the objective as follows,
\begin{equation*}
    L = \mathbb{E}_{\enc (\img),p,\varepsilon\sim N(0,1),t} \left [ \left \| \varepsilon _t-{f}_{\theta }(z_t,t,\enc(\img),p) \right \|_{2}^{2}  \right ] 
\end{equation*}
where ${f}_{\theta }$ represents a U-Net model tasked with predicting the noise $\varepsilon _t$ at each step of the process. The authors gather pairs of images \img, as shown in~\autoref{fig:ipx_pair}, differing only as described by the text instruction $p$. Therefore, the U-Net model learns to produce corresponding edited images based on the input image and text prompt from Gaussian noise. Importantly, \ipx applies Classifier-free guidance~\cite{classifier_free}, which makes the sampled images better correspond with the conditioning. Additionally, it employs a linear combination strategy to balance the influence of conditional and unconditional factors during model training. 

Here, we introduce the unique classifier-free guidance~\cite{classifier_free} design of \ipx~\cite{ipx}: To balance the effects of conditional and unconditional aspects during model training, the basic linear combination of the conditional and unconditional score estimates is designed as follows:  
\begin{equation*}
   \tilde{f}_{\theta }(z_t,c) = (1+s) {f}_{\theta }(z_t,c)-s{f}_{\theta }(z_t,\varnothing )
\end{equation*}
{in typical score estimation functions, $c$ is the condition, but} in our editing task, there are two conditions, \ie the original image \img and the edit instruction $p$. The modified score estimate is as follows, 
\begin{equation*}
\begin{aligned}
   \tilde{f}_{\theta }(z_t,\img,p) = &\; {f}_{\theta }(z_t,\varnothing ,\varnothing) 
    \\  +&\; s_I\cdot[{f}_{\theta }(z_t,\img,\varnothing)-{f}_{\theta }(z_t,\varnothing ,\varnothing )]
    \\  + &\; s_T\cdot[{f}_{\theta }(z_t,\img,p)-{f}_{\theta }(z_t,\img,\varnothing )]
\end{aligned}
\end{equation*}
Where the guidance scales $s_I$ and $s_T$ are hyper-parameters that adjust the relevance between generated samples and the input image and edit instruction, respectively.


\zztitle{\igc} It does not need pairs of images like \ipx, due to its unique fine-tuning paradigm. Their approach involves three steps. In the first step, it encodes the edit instruction with CLIP text~\cite{clip} and optimizes the embedding, making it close to the input image in the latent space. In the second step, it fine-tunes the generative diffusion model, \ie Stable Diffusion~\cite{latent_space}, with this embedding for improved image reconstruction. In the last step, it linearly interpolates the target and optimized text embeddings. This blended representation undergoes the diffusion process in their fine-tuned model, producing the final edited image.
Additionally, \igc freely integrates with various text-to-image diffusion models~\cite{imagic}. The researchers assessed their editing approach on both Imagen~\cite{imagen} and Stable Diffusion~\cite{latent_space}. {In our study, we employed the editing pipeline from version ``sd-v1-4-full-ema.ckpt'' of Stable Diffusion available on GitHub\footnote{https://github.com/justinpinkney/stable-diffusion/blob/main/notebooks/imagic.ipynb}, which is highly similar in structure and training data to the version ``sd-v1.5.ckpt'' used by \ipx.}

\zztitle{Text Edit Prompt} Brooks \el used $700$ captions from the LAION-Aesthetics dataset~\cite{schuhmann2022laion}, enhanced with manual editing instructions and output captions~\cite{ipx}. They fine-tuned a GPT-3 model~\cite{gpt3} for automated instruction generation. We evaluate using this augmented LAION-Aesthetics dataset. However, different models have entirely different styles of edit instructions, as shown in~\autoref{tab:edit_style}. The edit instruction for \ipx starts with a verb, essentially functioning as a command, whereas the edit instruction for \igc describes the image after the editing process to guide the model's editing. Brooks \el provide two styles of edit instructions based on the LAION-Aesthetics dataset~\cite{ipx}. However, the TedBench dataset has no corresponding edit instructions for the \ipx model. To address this, we generated the corresponding instruction styles using the GPT-4 API, as shown in~\autoref{tab:instGen}. Compared to fine-tuning with GPT-3 and data collecting, which costs thousands of dollars~\cite{git_ipx}, this approach costs 6.2 dollars while achieving identical results.

\section{Additional Results}
\label{app:add_exp}
\autoref{fig:edit_image} displays two examples: one is a target image with an image watermark, and the other with a text watermark. We directly edit these two images using the \ipx model. The corresponding results of these edits show that watermarks using images often become difficult to discern after being edited by a diffusion model, blending more into the original image. In contrast, text watermarks can be preserved. This may also contribute to the easier extraction of text watermarks. 

\autoref{fig:ipx_encode} demonstrates the varying degrees of visibility of text watermarks under editing by the \ipx and \igc models, throughout the optimization process, as encoder weight changes. For the \ipx model, the watermark remains inconspicuous at a higher weight value, \ie $\lambda=1/128$. In contrast, with the \igc model, the watermark in the edited images becomes progressively more distinct as $\lambda$ reaches $1/2$.

\autoref{fig:dist_acc_trend} illustrates the variation in accuracy of our three watermark authentication metrics—$\mathbf{D_{all}}$, $\mathbf{D_{word}}$, and $\mathbf{D_{letter}}$—for images before and after editing in the \ipx~\cite{ipx} and \igc~\cite{imagic} models, as the Semantic~\cite{clip} and Vision~\cite{vgg} distances increase. Overall, we have observed a consistent trend where an increase in distance leads to a decrease in accuracy. This phenomenon can be primarily attributed to the extent of modifications made to the image during the editing process. As the alterations become more substantial, they tend to have a significantly greater impact on the effectiveness of the watermarks we have implemented. 
\begin{figure}[t]
    \centering
    \begin{subfigure}{0.43\textwidth}
    \includegraphics[width=.95\linewidth]{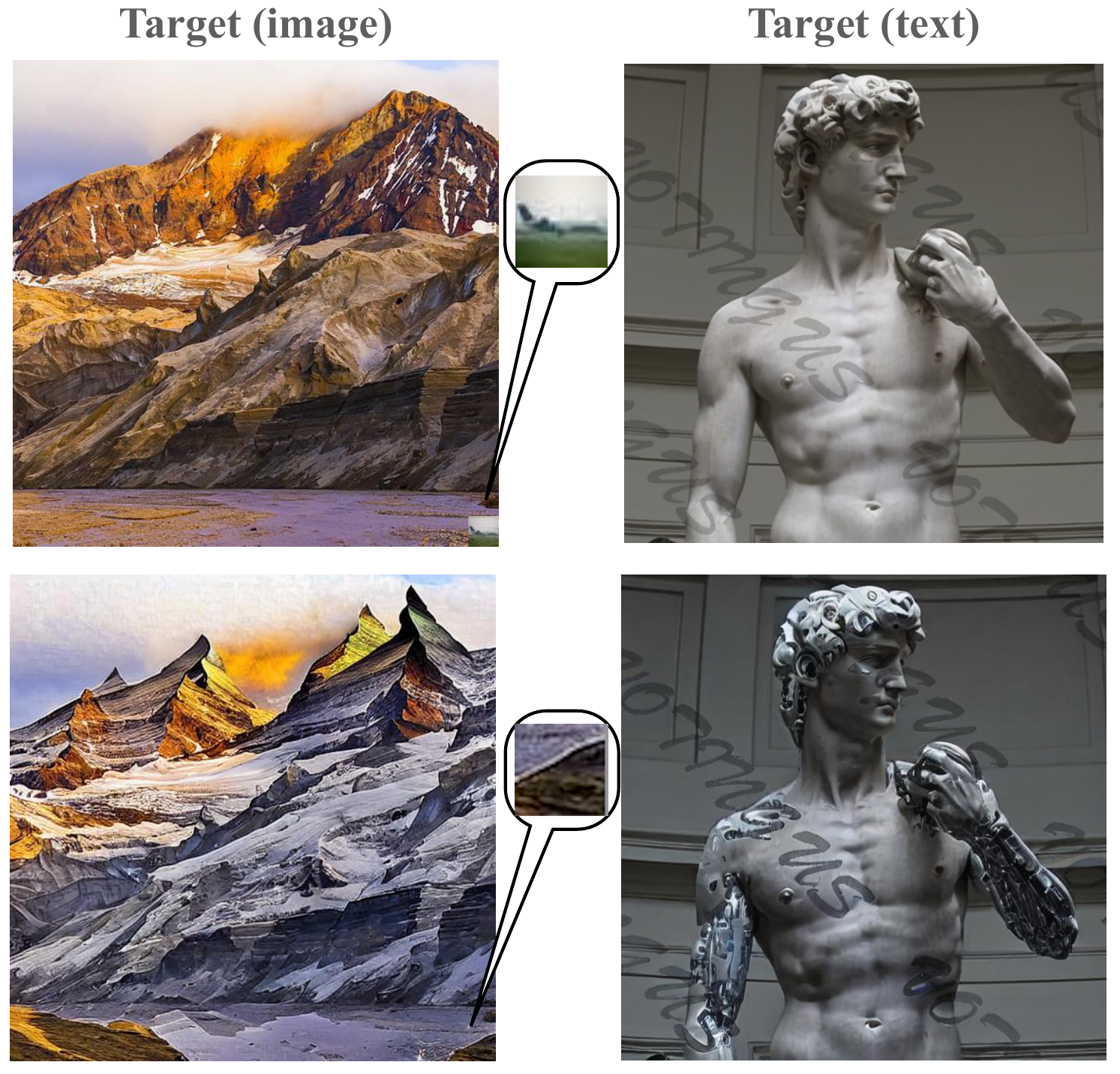}
    \end{subfigure}
    \caption{Examples showing some intuitions why image watermarks (left column) do not work while text watermarks (right column) do. The first row displays the target, which is a CIFAR10 image overlapped at the bottom right corner of the original image, and the text overlapped on an image. {The second row shows the corresponding edited images. The image watermark in the bottom right image does not retain much information about the watermark after editing. In contrast, the text watermark remains clearly visible.}}
    \label{fig:edit_image}
\end{figure}

\begin{figure*}[t]
    \centering
    \begin{subfigure}{0.95\textwidth}
        \centering
        \includegraphics[width=.95\linewidth]{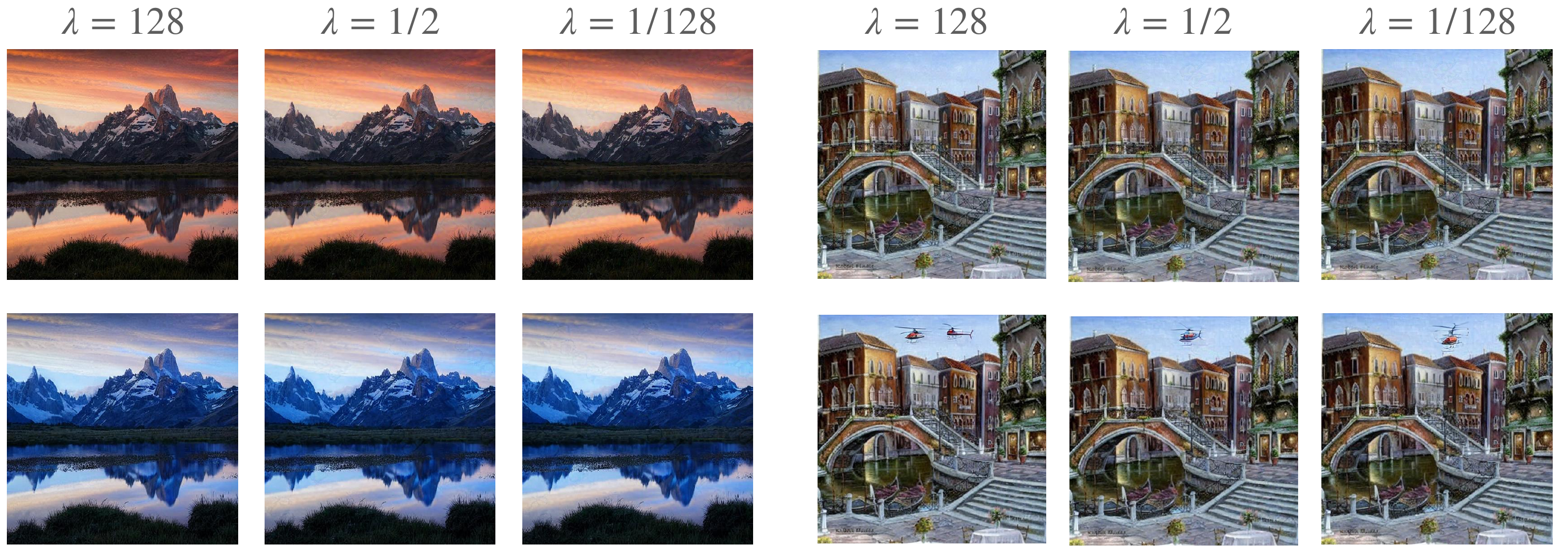}  
        \caption{Watermarking encoder weight $\lambda$ difference with \ipx model.}
    \end{subfigure}
    \begin{subfigure}{0.95\textwidth}
        \centering
        \includegraphics[width=.95\linewidth]{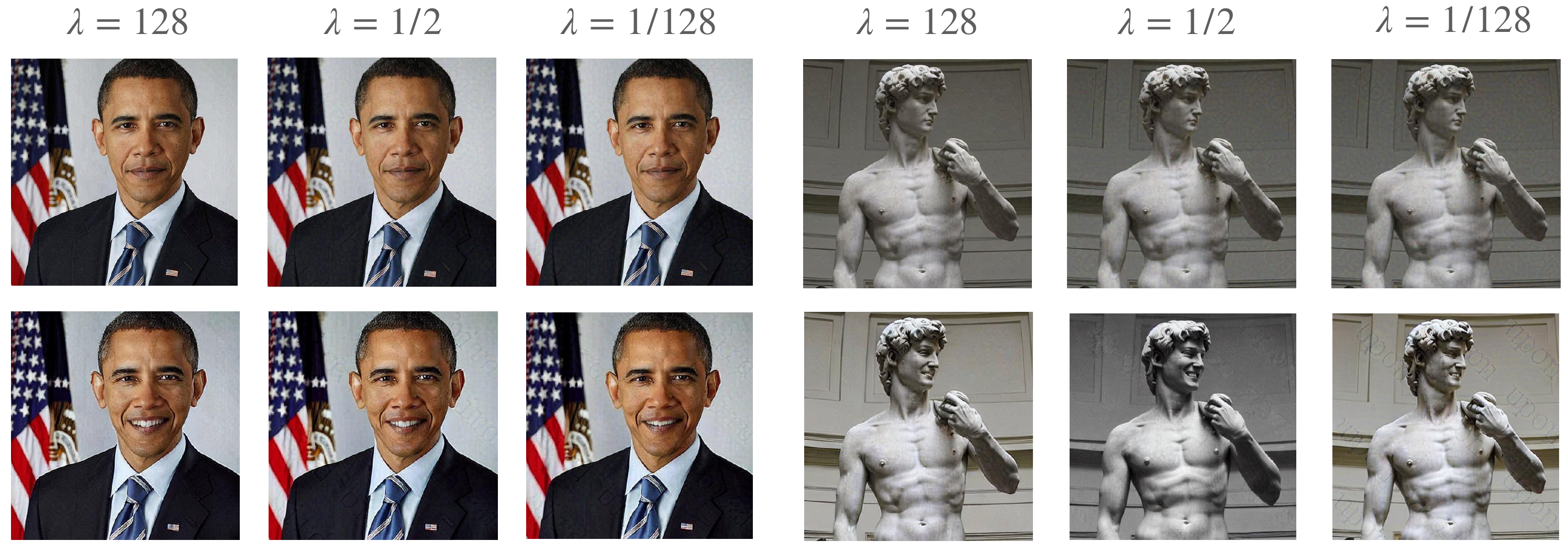}  
        \caption{Watermarking encoder weight $\lambda$ difference with \igc model.}
    \end{subfigure}
    \caption{The trend in encoder loss weight for both \ipx and \igc models, as observed in (a) and (b), reveals that at a minimal weight such as $128$, watermarks become entirely invisible post-edit. Conversely, at lower weights like $1/128$, watermarks become observable after editing. An optimal trade-off is found at a weight of $1/2$, where watermarks typically remain unseen post-edit and yet have a high probability of being extracted by \ext.}
    \label{fig:ipx_encode}
\end{figure*}

\begin{figure*}[t]
    \centering
    \begin{subfigure}{0.32\textwidth}
    \includegraphics[width=.95\linewidth]{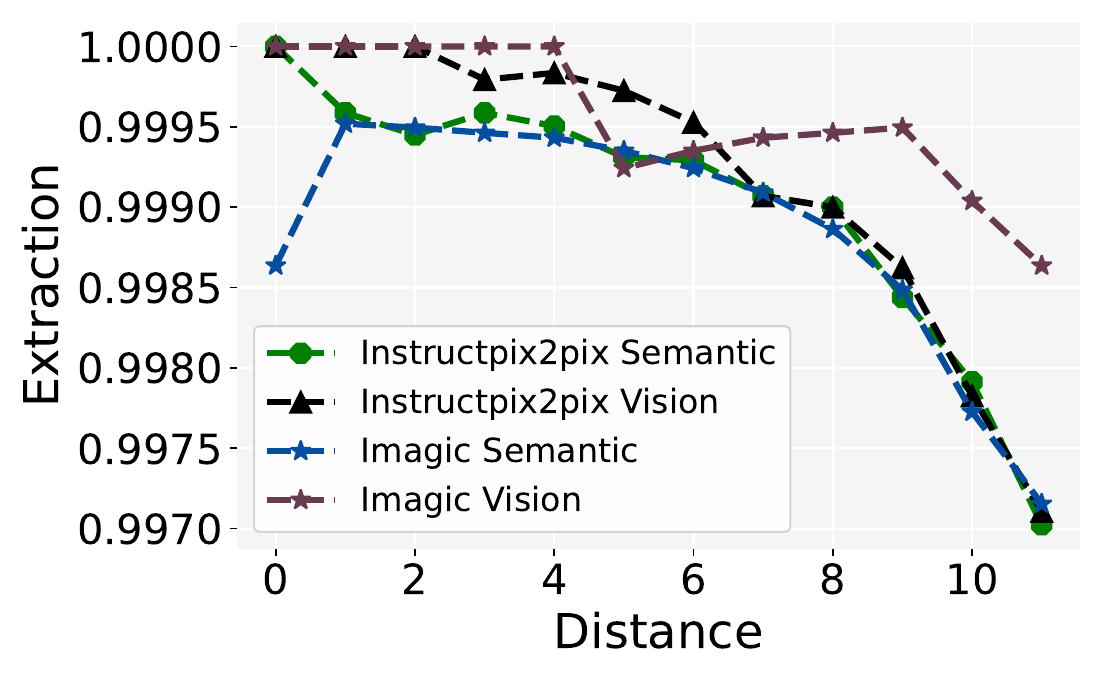}
    \caption{ $\mathbf{D_{letter}}$ accuracy}
    \end{subfigure}
    \begin{subfigure}{0.32\textwidth}
    \includegraphics[width=.95\linewidth]{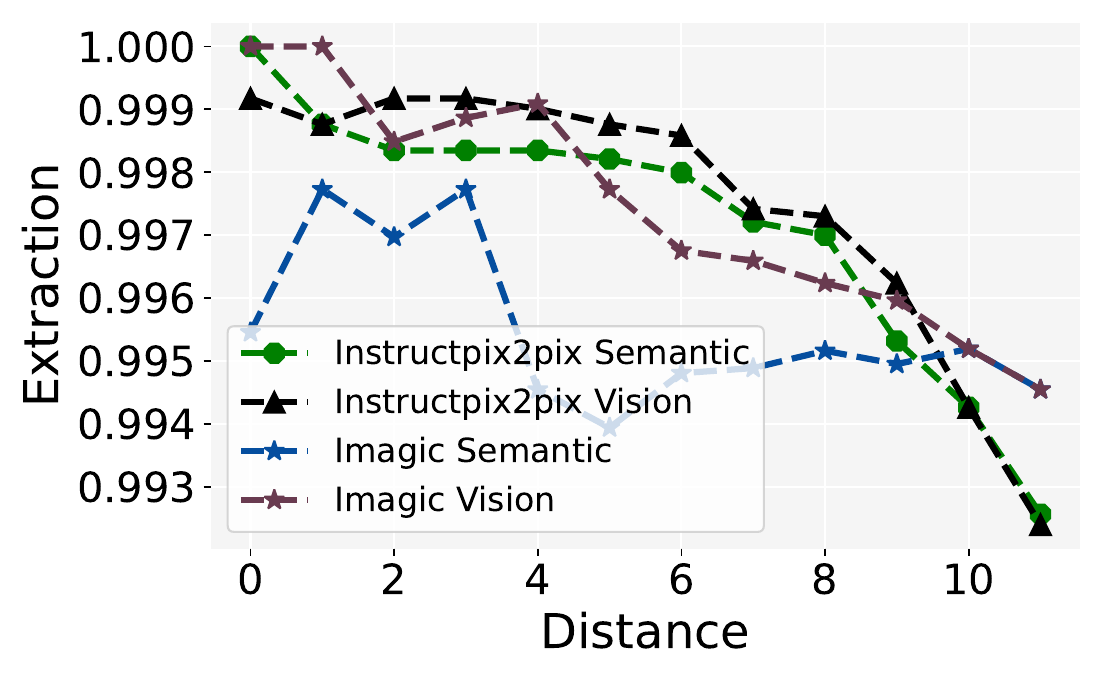}
    \caption{ $\mathbf{D_{word}}$ accuracy}
    \end{subfigure}
    \begin{subfigure}{0.32\textwidth}
     \includegraphics[width=.95\linewidth]{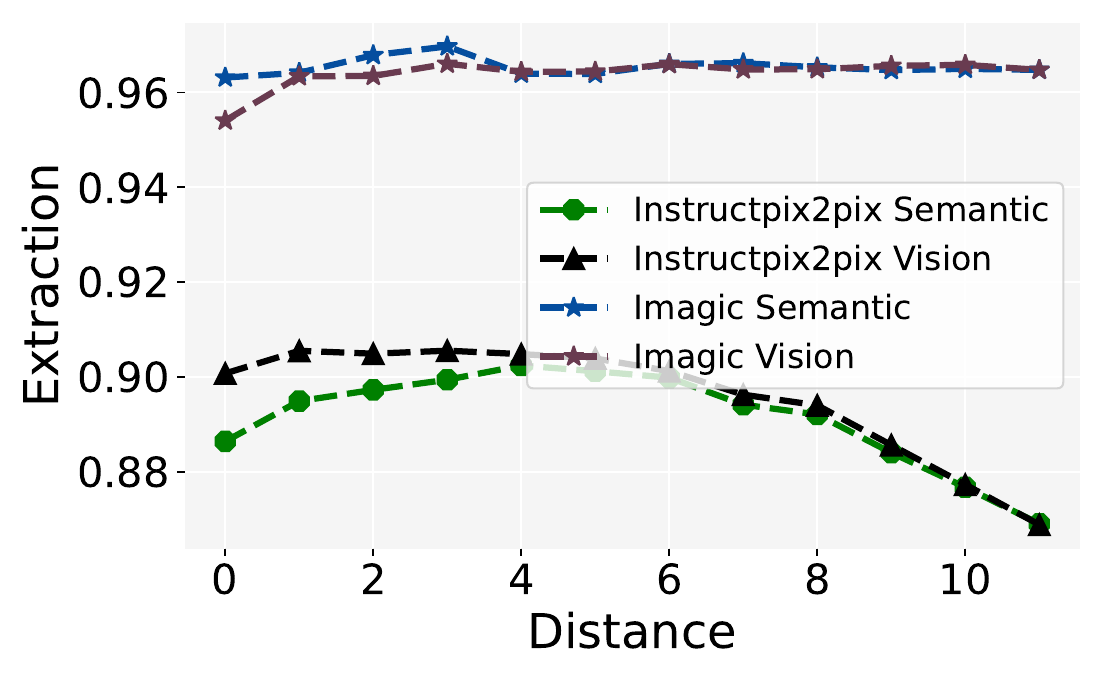}
    \caption{$\mathbf{D_{all}}$ accuracy}
    \end{subfigure}
    \caption{The watermark extraction accuracy diminishes with increasing edit distance. As edit distance often correlates with the extent of image modifications, extensive edits lead to reduced preservation of the original image's IP information, consequently diminishing IP infringement on the original image.}
    \label{fig:dist_acc_trend}
\end{figure*}

\end{document}